\documentclass[twocolumn]{aastex631}

\usepackage{multirow}
\usepackage{longtable}
\usepackage{comment}
\shorttitle{Precision Requirements for Robust Small Planet Classification}
\shortauthors{Cayman T. Unterborn \& Stephen Kane}

\begin{document}

\title{Precision Requirements for Robust Small Planet Classification}

\author[0000-0001-8991-3110]{Cayman T. Unterborn}
\affiliation{Space Sciences Department, Southwest Research Institute, San Antonio, TX, 78238}
\email{cayman.unterborn@swri.org}

\author[0000-0002-7084-0529]{Stephen R. Kane}
\affiliation{Department of Earth and Planetary Sciences, University of California, Riverside, CA 92521, USA}


\begin{abstract}

Classifying small exoplanets as rocky, volatile-rich, or iron-enriched is a prerequisite for efficient allocation of atmospheric characterization resources. The Nominally Rocky Planet Zone (NRPZ) provides a framework for this classification by comparing observed bulk densities against the expected density range for volatile-free rocky planets, calibrated from stellar abundance distributions. However, the efficacy of classifying a planet using the NRPZ is critically dependent on the precision of the underlying mass and radius measurements. Here, we perform an analytical propagation of observational uncertainties through the NRPZ classification framework for 37 candidate rocky exoplanets with radii below 2~$R_\oplus$ orbiting stars with visual magnitudes $<11$. Among the 37 planets, 12 already possess sufficiently precise mass and radius measurements to be confidently classified as Mini-Neptunes or Super-Mercuries. The remaining planets require approximately a $\sim$60\% reduction in the combined mass and radius uncertainties before they can be reliably classified. We find that  1--10$^3$ new observations are necessary to achieve these lower uncertainties, and these are likely lower limits. We argue that, while possible, this level of intense observation is likely not feasible for many of these planets and that mass-radius models alone are not likely to allow us to classify all but those planets with outlier densities. We therefore recommend a staged strategy in which improved mass and radius measurements are used to identify and prioritize ambiguous systems, followed (where feasible) by atmospheric observations with JWST or future facilities as a complementary test of whether selected planets retain substantial atmospheres.

\end{abstract}



\section{Introduction and Science Motivation}




The classification of small exoplanets represents one of the most consequential challenges in contemporary exoplanetary science. Over the past two decades, transit and radial velocity surveys have revealed that planets with radii between 1--2~$R_\oplus$ are among the most abundant outcomes of planet formation \citep{howard2010a,petigura2013a,fressin2013,dressing2015a}, yet their bulk compositions remain remarkably difficult to constrain from mass and radius measurements alone \citep{rogers2015a,dorn2015,unterborn2023}. Resolving this ambiguity is essential for several interconnected reasons. First, determining whether a given planet is rocky or possesses a significant volatile envelope is a prerequisite for assessing its potential habitability, since the presence or absence of an atmosphere fundamentally governs surface conditions and the plausibility of liquid water \citep{kopparapu2013a,kane2021e,bean2017}. Second, robust compositional classification enables the comparative study of terrestrial planet evolution across diverse stellar environments, providing empirical tests of models that connect interior structure, outgassing history, and long-term climate stability to the divergent fates of worlds like Earth and Venus \citep{kane2021d,kane2024b}. Third, at the population level, the ability to reliably distinguish rocky planets from those retaining primordial or secondary envelopes directly constrains the physical mechanisms responsible for sculpting the observed radius valley \citep{fulton2018b, vaneylen2018b}, including photoevaporation, core-powered mass loss, and water-world formation pathways \citep{owen2017c,lopez2014,fulton2017,luque2022b}. Finally, the efficient allocation of atmospheric characterization resources with the James Webb Space Telescope \citep[JWST;][]{gardner2006} and future facilities such as the Habitable Worlds Observatory \citep[HWO;][]{NationalAcademies2021} requires accurate target prioritization, since the scientific return of spectroscopic follow-up depends critically on the nature of the target \citep{kempton2018,greene2016,batalha2014}. In all of these contexts, the fidelity of compositional classification is limited by the precision of the underlying mass and radius measurements, and the current state of these uncertainties for the majority of known small exoplanets remains insufficient for definitive determination.

One such classification scheme is that of the Nominally Rocky Planet Zone \citep[NRPZ, ][]{unterborn2023}. The NRPZ is calculated under the assumption that the stellar abundances of the major rocky planet-building elements Mg, Si, and Fe in the Hypatia Catalog of FGK stars represent the same broad distribution as that found in planets. This assumption is grounded in the fact that the relative condensation temperatures of these elements are insensitive to the bulk abundances of these elements and that the relative ratios of these elements (e.g., Fe/Mg) are unlikely to change due to planet formation \citep{bond2010a, unterborn2017}, and is supported by multiple studies \citep[e.g.,][]{Thiabaud2015, schulze2021, Adibekyan2021, unterborn2023, Adibekyan2024}. Under the assumption that planetary compositions broadly reflect those of their host stars, \citet{unterborn2023} stoichiometrically determined the resulting 3$\sigma$ range of Fe-core mass fractions and calculated the resulting 3$\sigma$ range planet radii for a given mass using the ExoPlex mass-radius-composition calculator \citep{unterborn2018a, unterborn2018b, Unterborn2019, unterborn2023}. By comparing the measured mass, radius, and their respective uncertainties to the location of the NRPZ, \citet{unterborn2023} showed that we can assign a probability that a planet is consistent with being a rocky planet without a significant volatile-rich atmosphere. While the one-to-one assumption that stellar abundances reflect planet abundances is debated \citep[e.g., ][]{Plotnykov2020, Adibekyan2021, Liu2023, Adibekyan2024}, the NRPZ is meant to be a first-order diagnostic of whether a planet is consistent with a rocky composition while taking into account the precision in mass and radius uncertainties. Further discussion of the NRPZ is located in \S\ref{sec:NRPZ}.

Such precision is important not only for identifying the best targets for atmospheric characterization and potential habitability studies, but also for placing terrestrial planets into a broader evolutionary context. In particular, accurate classification of planets as rocky, volatile-rich, or iron-enriched is essential for population-level studies of the rocky/non-rocky transition, the radius valley, and the relative roles of formation environment, volatile accretion, and atmospheric loss in shaping the observed small-planet population \citep{rogers2015a,fulton2017,fulton2018b,owen2017c,luque2022b,unterborn2023}. In that sense, improved radius precision directly strengthens both the classification of individual planets and the statistical interpretation of terrestrial planet evolution across exoplanet populations \citep{berger2018c,christiansen2025}.

In this paper, we explore the precision needed in mass and radius uncertainties that are required to classify planets orbiting the brightest targets in our current sample of exoplanets. In section \S\ref{sec:NRPZ}, we describe the NRPZ in more detail. Section \S\ref{sec:sample} discusses our sample of small, cool exoplanets that orbit bright targets. In section \S\ref{sec:error_prop}, we outline our methodology for simulating new mass and radius measurements and determine the required uncertainties for classification of these planets. In section \S\ref{sec:best_uncertainties}, we outline the minimum uncertainties in mass and radius needed and the observational time needed to classify planets as rocky or not rocky. In section \S\ref{sec:feasibility} we discuss the observational feasibility of obtaining these uncertainties. Finally in section \S\ref{sec:discussion} we discuss the limitations of our model and future observational priorities determining whether intermediate density planets are, in fact, rocky. 

\section{The Nominally Rocky Planet Zone}
\label{sec:NRPZ}
Mass-radius models have been the gold standard for classifying and characterizing exoplanets for over two decades \citep{valencia2006, valencia2007a, valencia2007b, seager2007, dorn2015, zeng2016a, zeng2016b, unterborn2016,unterborn2018a, zeng2019, Magrathea_2022, unterborn2023}. The traditional method for classifying a newly discovered exoplanet is to compare the observed mass and radius to multiple single iso-compositional mass-radius curves, with the nearest curve often being used as the inferred composition. This method is known to be fraught with degeneracy for intermediate density exoplanets with multiple compositions often providing the same density \citep{Seager2003, dorn2015, unterborn2023}. For example, the TRAPPIST-1 exoplanets have the lowest relative uncertainties in mass and radius, averaging $\sigma_M$/M = 3.9\% and $\sigma_R$/R = 1.4\%. Even at these extremely low uncertainties, the TRAPPIST-1 b, d and g have low enough measured densities \citep{gillon2017a} to be equally likely to have Fe-core mass fractions less than the Earth or an Earth-like Fe-core mass fraction with $<$40 mass percent water on their surfaces \citep{unterborn2018b}. This degeneracy is exacerbated as uncertainties in mass and radius increase, with even more iso-compositional mass-radius curves crossing within the observed values. Definitively classifying an intermediate-density exoplanet via mass-radius curves alone is quite difficult. 
\begin{figure*}
    \centering
    \includegraphics[width=\linewidth]{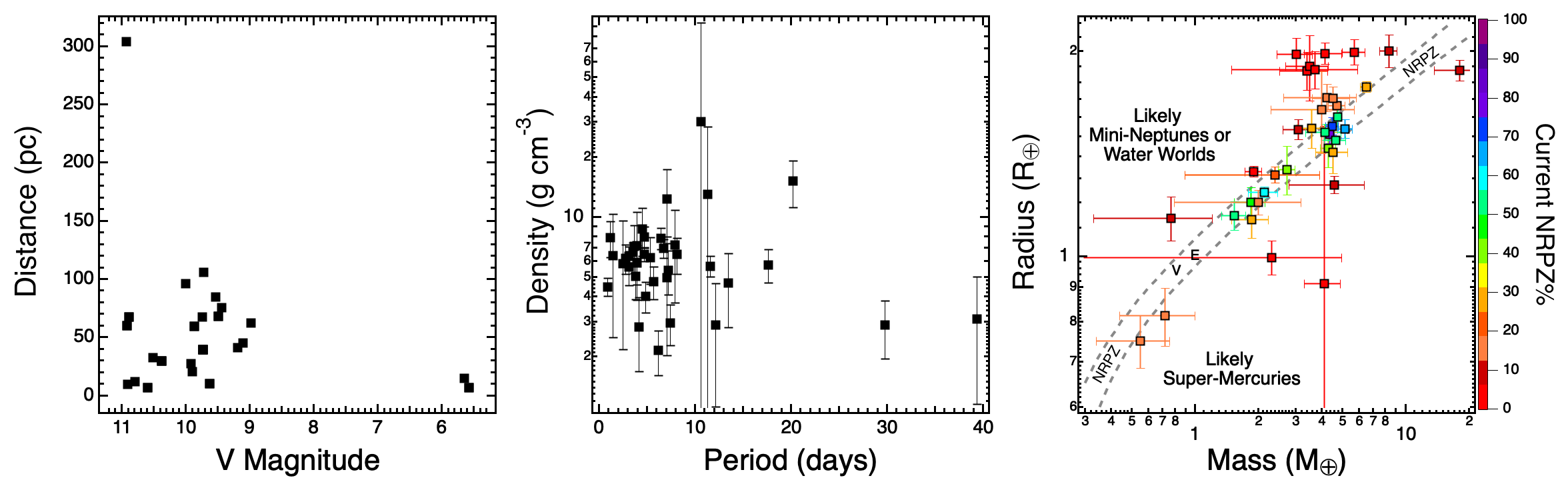}
    \caption{Distance and visual magnitude (\textbf{left}), density and period (\textbf{center}) and mass and radius color coded by their current NRPZ\% (\textbf{right)} for our sample of 37 exoplanets with visual magnitudes below 11 and equilibrium temperatures below 1300 K. The Earth and Venus are included for reference.} Data for all planets taken from the Exoplanet Archive \citep{Exo_archive} are shown in Table \ref{tab:sample}.
    \label{fig:sample}
\end{figure*}

To remedy this dependency on individual mass-radius curves being used for classification, \citet{unterborn2023} created the Nominally Rocky Planet Zone (NRPZ). The NRPZ defines the 3$\sigma$ (99.7\%) range of bulk densities expected for volatile-free, rocky planets without a substantial atmosphere as a function of planetary radius. This density range is computed using the ExoPlex interior structure code, which models the internal structure of rocky planets across the full range of Fe/Mg/Si abundance ratios observed in FGK host stars from the Hypatia Catalog \citep{hinkel2014}. The resulting NRPZ spans a density ratio of approximately 0.85--1.10 relative to an Earth-composition reference model, corresponding to iron core mass fractions (CMFs) of roughly 0.25--0.45. The NRPZ defines the range of all densities consistent with solid, atmosphere-free, planet compositions based on the range expected from stellar abundance data. Planets with mass and radius values that fall within the NRPZ are therefore consistent with being rocky without a significant atmosphere and having a composition commensurate with stellar abundance distributions. Planets whose masses and radii fall outside of the NRPZ would therefore require other compositional factors to be at play to explain their higher or lower densities. Planets with density lower than the NRPZ would require some amount of surface volatiles such as an extended atmosphere or water layers and can thus be classified as Mini-Neptunes or water worlds, depending on the species of surface volatile. Planets with density greater than the NRPZ would require an enrichment of Fe relative to the other rock-building elements above that expected from stellar abundance data, and would thus be classified as Super-Mercuries. 

For a given planet, we can define its probability of being within the NRPZ (NRPZ\%) as the fraction of its measured mass-radius distribution, including uncertainties, that fall within the NRPZ. A planet's NRPZ\% is therefore: 
\begin{equation}
    \rm{NRPZ}\%  = 100\times\frac{\Sigma_{i}^{N}\left(\rm{M}_i, \rm{R}_i\right)_{\rm{within \ NRPZ}}}{N}. 
    \label{eq:NRPZ}
\end{equation}
which is simply the total number of individual mass-radius pairs that fall within the NRPZ as defined by equations 25 and 28 of \citet{unterborn2023}, divided by the total number of mass-radius pairs, $N$, assuming an independent bivariate normal distribution in both mass and radius based on their observed values and their respective uncertainties. For example, TRAPPIST-1 b has a mass of 1.374$\pm$0.069 M$_\oplus$, a radius of 1.116$\pm$0.014 R$_\oplus$, and a density of 5.45$\pm$0.34 g cm$^{-3}$ \citep{agol2021}, which yields an NRPZ\% of 98\% for $N=10^3$ \citep{unterborn2023} . In contrast, GJ 3929 b has a mass similar to TRAPPIST-1 b at 1.75$\pm$0.44 M$_\oplus$, but a smaller radius at 1.09$\pm$0.04 R$_\oplus$, and thus a larger density of 7.45 $\pm$ 2.04 g cm$^{-3}$ \citep{beard2022a}. This higher density places it outside the NRPZ, yielding an NRPZ\% for GJ 3929 b of only 22\% for $N=10^3$. Although GJ 3929 b has much larger relative bulk density uncertainties than TRAPPIST-1 b ($\sim27\%$ versus $\sim8\%$), its substantially higher mean density places it well outside the NRPZ. The NRPZ method would therefore classify GJ 3929 b as a super-Mercury at roughly the $\sim1\sigma$ confidence level. 

In general, the NRPZ produces two classifications, planets with NRPZ\% $<$ 25\% are considered ``not rocky,'' while those with NRPZ\% $>$ 75\% can be considered ``rocky'' without significant surface volatile layers at the $\sim1\sigma$ confidence level. TRAPPIST-1 b and GJ 3929 b show the utility of the NRPZ in classifying exoplanets in a broad context. TRAPPIST-1 with its average mass and radius coupled with very small uncertainties in both is of intermediate density and is definitively within the NRPZ. GJ 3929 b, on the other hand, shows how planets with larger uncertainties can be classified if their average mass, radius and density place them far enough outside of the NRPZ to begin with. For intermediate density exoplanets then, small uncertainties are required for definitive classification, while the NRPZ is still capable of identifying and classifying non-rocky exoplanets even if they have large uncertainties. Unfortunately, the majority of intermediate density exoplanets have uncertainties in mass and radius too large to definitively classify them as either rocky or not rocky, with \citet{unterborn2023} only identifying 9 planets that are able to be characterized as rocky, five of which are the TRAPPIST-1 exoplanets.


\section{Sample of Exoplanets}
\label{sec:sample}
\begin{figure*}
    \centering
    \includegraphics[width=\linewidth]{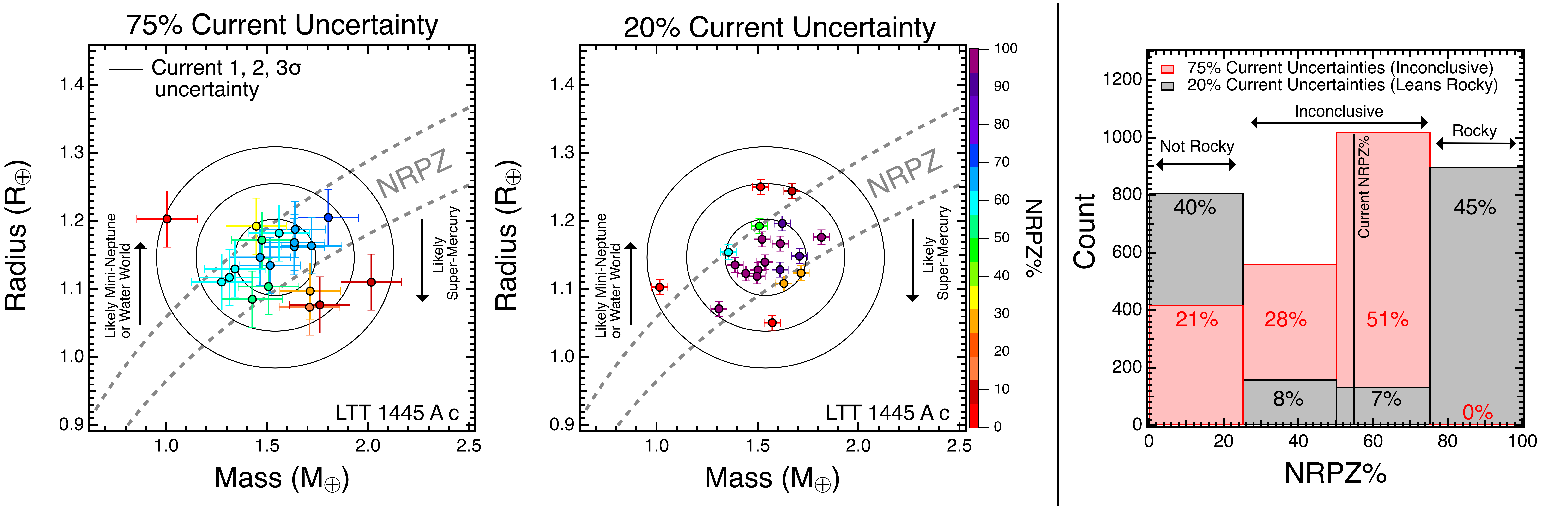}
    \caption{The current 1, 2 and 3$\sigma$ error ellipses for LTT 1445 A c, with 20 of 2000 randomly selected average mass and radii assuming mass and radius uncertainties at 75\% (\textbf{left}) and 20\% (\textbf{center}) of the current values ($\sigma_M$ = 0.2 M$_\oplus$, $\sigma_R$ = 0.055 R$_\oplus$). Colors represent NRPZ\% of an individual model. \textbf{Right:} Resulting histogram of NRPZ\% for all 2000 average masses and radii selected. Error ellipses are used to determine the new values of $M_{avg}$ and $R_{avg}$ because new measurements may change these values. Lower uncertainties are then assigned to determine the new NRPZ distribution. At 75\% uncertainties, only 21\% of models are classifiable with more than half having an NRPZ similar to the current value of $\sim55\%$. At 20\% of the current uncertainties, 85\% of models would be considered classifiable with the exact classification of rocky or not rocky depending on the average mass and radius achieved upon a new measurement.}
    \label{fig:example}
\end{figure*}

The major source in future reductions in the uncertainties in mass and radius are likely to be achieved by Extreme Precision Radial Velocity (EPRV) instruments. As EPRV follow-up observations are most sensitive to brightness, we limit ourselves to only those planets with visual magnitudes below 11. To maximize the likelihood that these planets could be rocky, we limit ourselves to those with radii $<$2 R$_\oplus$, putting us near or below the supposed ``radius-valley'' of \citet{fulton2017}. According to the NASA Exoplanet archive \citep{Exo_archive} as of March, 2026, 64 planets have both mass and radius measurements available and meet these criteria. Finally, we wish to consider only those planets that are likely solid and thus we limit ourselves to those planets with equilibrium temperatures below the dry melting temperature of peridotite \citep[$\sim$1300 K,][]{Katz_2003}. While lava worlds are still possible in our sample due to either their age or the presence of a greenhouse atmosphere, these are unlikely to be characterized as molten from mass-radius models alone due to underlying degeneracies. Only follow-up observations such as atmospheric characterization are likely to definitively classify them as lava worlds, which we discuss later. Of the 64 planets, 37 have equilibrium temperatures below 1300 K and represent our sample of exoplanets for this work (Figure \ref{fig:sample}, Table \ref{tab:sample}).

The 37 planets in our sample span masses and radii between 0.5--18 M$\oplus$ and 0.75--2 R$_\oplus$, respectively, and thus densities between 2--30 g cm$^{-3}$. The average relative uncertainties for these planets in $\sigma_M$/M and $\sigma_R$/R of 25\% and 7\%, respectively. These planets have NRPZ probabilities of between 0--80\% when utilizing their current masses, radii and their respective uncertainties, with an average NRPZ\% of 26\%. They lie between $\sim6$--300 pc away, with an average of $\sim$50 pc (Figure \ref{fig:sample}, left). They have orbital periods between 0.9--39 days, with an average of 8 days (Figure \ref{fig:sample}, center). Although many of the stars a relatively close ($< 100$~pc), their brightnesses vary considerably due to many of the host stars being M dwarfs. Those targets most suitable to EPRV are the F--G--K stars with brightnesses of $V < 10$.

\begin{table*}
    \centering
    
    \caption{Sample of Exoplanets available mass and radius measurements that have Visual Magnitude $<$ 11 and Equilibrium Temperature $<$ 1300 K. All data with the exception of NRPZ\% derived from the Exoplanet Archive \citep{Exo_archive}.}
    \begin{tabular}{lccccccc}
        & Current &  Stellar& Visual & Distance & Period & Radius & Mass \\   
        Planet & NRPZ\% & T$\rm{_{eff}}$ (K) & Magnitude& (pc) & (days) & (R$_\oplus$) & (M$_\oplus$)   \\ 
        \hline
        GJ 357 b & 47 & 3505 & 10.91 & 9.44 & 3.93 & 1.20$\pm$0.06 (5.0\%)\tablenotemark{a}  & 1.84$\pm$0.31 (16.8\%) \\ 
        GJ 806 b & 5 & 3600 & 10.79 & 12.04 & 0.93 & 1.33$\pm$0.02 (1.7\%) & 1.90$\pm$0.17 (8.9\%) \\ 
        GJ 9827 b & 41 & 4236 & 10.37 & 29.66 & 1.21 & 1.44$\pm$0.09 (6.3\%) & 4.28$\pm$0.35 (8.2\%) \\ 
        GJ 9827 c & 29 & 4236 & 10.37 & 29.66 & 3.65 & 1.13$\pm$0.07 (6.2\%) & 1.86$\pm$0.37 (19.9\%) \\ 
        GJ 9827 d & 0 & 4236 & 10.37 & 29.66 & 6.2 & 1.98$\pm$0.11 (5.6\%) & 3.02$\pm$0.58 (19.2\%) \\ 
        HD 15337 b & 31 & 5131 & 9.1 & 44.82 & 4.76 & 1.77$\pm$0.03 (1.8\%) & 6.52$\pm$0.41 (6.3\%) \\ 
        HD 23472 b & 11 & 4684 & 9.73 & 39.03 & 17.67 & 2.00$\pm$0.11 (5.5\%) & 8.32$\pm$0.78 (9.4\%) \\ 
        HD 23472 c & 0 & 4684 & 9.73 & 39.03 & 29.8 & 1.87$\pm$0.12 (6.4\%) & 3.41$\pm$0.88 (25.8\%) \\ 
        HD 23472 d & 15 & 4684 & 9.73 & 39.03 & 3.98 & 0.75$\pm$0.07 (8.9\%) & 0.55$\pm$0.21 (38.2\%) \\ 
        HD 23472 e & 17 & 4684 & 9.73 & 39.03 & 7.91 & 0.82$\pm$0.08 (9.8\%) & 0.72$\pm$0.28 (38.9\%) \\ 
        HD 23472 f & 11 & 4684 & 9.73 & 39.03 & 12.16 & 1.14$\pm$0.08 (7.4\%) & 0.77$\pm$0.44 (57.1\%) \\ 
        HD 136352 b & 17 & 5664 & 5.65 & 14.68 & 11.58 & 1.66$\pm$0.04 (2.6\%) & 4.72$\pm$0.42 (8.9\%) \\ 
        HD 219134 b & 56 & 4699 & 5.57 & 6.53 & 3.09 & 1.60$\pm$0.06 (3.4\%) & 4.74$\pm$0.19 (4.0\%) \\ 
        HD 219134 c & 80 & 4699 & 5.57 & 6.53 & 6.76 & 1.51$\pm$0.05 (3.1\%) & 4.36$\pm$0.22 (5.0\%) \\ 
        HD 224018 b & 5 & 5784 & 9.72 & 105.48 & 10.64 & 0.91$\pm$0.64 (70.3\%) & 4.10$\pm$0.80 (19.5\%) \\ 
        HD 235088 b & 0 & 5075 & 9.19 & 41.17 & 7.43 & 1.98$\pm$0.07 (3.6\%) & 4.15$\pm$0.83 (20.0\%) \\ 
        HD 260655 b & 61 & 3803 & 9.63 & 10.01 & 2.77 & 1.24$\pm$0.02 (1.9\%) & 2.14$\pm$0.34 (15.9\%) \\ 
        HD 260655 c & 12 & 3803 & 9.63 & 10.01 & 5.71 & 1.53$\pm$0.05 (3.3\%) & 3.09$\pm$0.48 (15.5\%) \\ 
        HIP 29442 c & 74 & 5289 & 9.49 & 68.19 & 3.54 & 1.55$\pm$0.05 (2.9\%) & 4.50$\pm$0.32 (7.1\%) \\ 
        HIP 29442 d & 65 & 5289 & 9.49 & 68.19 & 6.43 & 1.54$\pm$0.05 (3.2\%) & 5.14$\pm$0.41 (8.0\%) \\ 
        K2-233 b & 21 & 4796 & 10.88 & 67.51 & 2.47 & 1.32$\pm$0.04 (2.7\%) & 2.40$\pm$1.50 (62.5\%) \\ 
        K2-233 c & 7 & 4796 & 10.88 & 67.51 & 7.06 & 1.27$\pm$0.04 (3.0\%) & 4.60$\pm$1.80 (39.1\%) \\ 
        Kepler-65 d & 55 & 6211 & 10.92 & 303.73 & 8.13 & 1.52$\pm$0.04 (2.6\%) & 4.14$\pm$0.79 (19.1\%) \\ 
        Kepler-93 b & 53 & 5669 & 10 & 95.91 & 4.73 & 1.48$\pm$0.02 (1.3\%) & 4.66$\pm$0.53 (11.4\%) \\ 
        LTT 1445 A b & 38 & 3562 & 10.59 & 6.87 & 5.36 & 1.34$\pm$0.11 (8.2\%) & 2.73$\pm$0.25 (9.2\%) \\ 
        LTT 1445 A c & 53 & 3340 & 10.59 & 6.87 & 3.12 & 1.15$\pm$0.06 (4.8\%) & 1.54$\pm$0.20 (13.0\%) \\ 
        TOI-260 b & 15 & 4026 & 9.9 & 20.19 & 13.48 & 1.71$\pm$0.08 (4.7\%) & 4.23$\pm$1.60 (37.8\%) \\ 
        TOI-286 b & 28 & 5152 & 9.87 & 59.23 & 4.51 & 1.42$\pm$0.10 (7.0\%) & 4.53$\pm$0.78 (17.2\%) \\ 
        TOI-286 c & 4 & 5152 & 9.87 & 59.23 & 39.36 & 1.88$\pm$0.12 (6.4\%) & 3.72$\pm$2.22 (59.7\%) \\ 
        TOI-512 b & 29 & 5277 & 9.74 & 67.4 & 7.19 & 1.54$\pm$0.10 (6.5\%) & 3.57$\pm$0.53 (14.8\%) \\ 
        TOI-733 b & 0 & 5585 & 9.44 & 75.21 & 4.88 & 1.99$\pm$0.09 (4.3\%) & 5.72$\pm$0.70 (12.2\%) \\ 
        TOI-836 b & 14 & 4552 & 9.92 & 27.5 & 3.82 & 1.70$\pm$0.07 (3.9\%) & 4.53$\pm$0.92 (20.3\%) \\ 
        TOI-1136 b & 3 & 5770 & 9.53 & 84.54 & 4.17 & 1.90$\pm$0.21 (11.1\%) & 3.50$\pm$0.80 (22.9\%) \\ 
        TOI-1411 b & 18 & 4115 & 10.51 & 32.48 & 1.45 & 1.20$\pm$0.05 (4.1\%) & 2.00$\pm$1.20 (60.0\%) \\ 
        TOI-1799 b & 16 & 5690 & 8.98 & 62.13 & 7.09 & 1.64$\pm$0.23 (13.8\%) & 4.00$\pm$1.70 (42.5\%) \\ 
        TOI-2322 b & 4 & 4664 & 10.91 & 59.86 & 11.31 & 0.99$\pm$0.06 (5.7\%) & 2.31$\pm$2.67 (115.6\%) \\ 
        TOI-2322 c & 10 & 4664 & 10.91 & 59.86 & 20.23 & 1.87$\pm$0.07 (3.5\%) & 18.10$\pm$4.34 (24.0\%) \\ 
    \end{tabular}
    \tablenotetext{a}{Relative uncertainty in radius or mass, $100*\sigma_X/X$}
    \label{tab:sample}
\end{table*}

\section{Mass and Radius Error Propagation}
\label{sec:error_prop}

To determine the minimum mass and radius uncertainties needed to classify these planets as either rocky or not rocky, we treat the reported mass and radius values and their published uncertainties as Gaussian posterior summaries for the unknown latent planet properties. We therefore draw 2000 possible average mass ($M_{avg}$) and radius ($R_{avg}$) pairs from independent normal distributions centered on the reported values with widths equal to the published $1\sigma$ uncertainties. These are draws of possible latent values rather than posterior-predictive repeat observations. The additional $\sqrt{2}$ broadening discussed by \citet{shy2022} applies when generating a new noisy observation conditional on an already noisy measurement and is therefore not applied to these latent-value draws. We then independently assign each $M_{avg}$ and $R_{avg}$ value a given mass and radius uncertainty. We then assume that this new set of $M_{avg}$ and $R_{avg}$ along with their reduced $\sigma_M$ and $\sigma_R$, represents a new measurement and we calculate the resulting NRPZ\% for this measurement using  equation \ref{eq:NRPZ}. To calculate NRPZ\%, we assume that $M_{avg}$, $R_{avg}$, $\sigma_M$ and $\sigma_R$ represent a multivariate normal distribution and calculate $N=2000$ mass-radius pairs within this distribution. We then repeat this process for all 2000 $M_{avg}$ and $R_{avg}$ pairs, which produces a distribution of 4$\times10^6$ NRPZ\% values for a single set of $\sigma_M$ and $\sigma_R$ values. We calculate new values of $M_{avg}$ and $R_{avg}$ for each $\sigma_M$ and $\sigma_R$ pair. For this work, we vary $\sigma_M$ and $\sigma_R$ independently, sampling on a 40x40 grid of reduced uncertainties in mass and radius spanning 100\% and $\sim4\%$ of their current values listed in Table \ref{tab:sample}.

By examining how our choice of $\sigma_M$ and $\sigma_R$ changes the distribution of NRPZ\%, we can understand how smaller uncertainties, along with changes in $M_{avg}$ and $R_{avg}$, affect our ability to classify a planet as rocky or not rocky. To determine whether $\sigma_M$ and $\sigma_R$ are sufficiently small enough to definitively classify a planet as rocky, we require that 75\% of all models must fall above an NRPZ\% value of 75\%. Similarly, we consider a planet as having sufficient uncertainties to define it as not rocky, that is outside of the NRPZ, when 75\% of all models fall below an NRPZ\% value of 25\%. Finally, models where the majority of the NRPZ\% distribution falls between 25\% and 75\%, are considered inconclusive, and unable to be classified.

If a planet cannot be classified as rocky or not rocky for a given set of reduced uncertainties, two possibilities arise: the majority of models fall within the inconclusive range of NRPZ\% (25--75\%) or a bimodal distribution is produced, with some large fraction of models in both the rocky and not rocky bins. We still consider this latter case, where uncertainties are small enough such that the sum of the rocky and not rocky bins reaches above 75\%, to be \textit{classifiable}. In this case, those models where the largest fraction of models falls within the NRPZ\% $\geq$75\% bin or NRPZ\% $\leq$25\% bin, we define as either leans rocky (LR) or leans not-rocky (LNR), respectively. Because our methodology assumes that $M_{avg}$ and $R_{avg}$ will change with any new EPRV or photometry measurements, the exact classification of a LR or LNR exoplanet will be determined by the newly acquired $M_{avg}$ and $R_{avg}$ values and will have sufficient uncertainty to classify them as such. 

For example, we follow this procedure for planet LTT 1445 A c (Figure \ref{fig:example}) which has a mass and radius of 1.54$\pm$0.25 M$_\oplus$ and 1.147$\pm$0.11 R$_\oplus$, respectively. Assuming these currently reported values of $M_{avg}$, $R_{avg}$ $\sigma_M$ and $\sigma_R$, LTT 1445 A c has a current NRPZ\% of $\sim$55\% following Equation \ref{eq:NRPZ}. Following the methodology described above, we show the results assuming a 75\% and 20\% reduction in both the current mass and radius uncertainties simultaneously, although in our results we vary both independently to test the sensitivity of both uncertainties in our analysis. At 75\% uncertainties, nearly half of our sample have $M_{avg}$ and $R_{avg}$ values that fall within the NRPZ, but these high uncertainties produce an NRPZ\% distribution that falls mainly between NRPZ\% = 25--75\% (Figure \ref{fig:example}). This means 79\% of all models produce an inconclusive classification. As the uncertainty and mass and radius decreases to 20\%, the fraction of $M_{avg}$ and $R_{avg}$ values within the NRPZ stays roughly the same, however now the NRPZ\% distribution shifts such that 85\% of all models fall within the not rocky (40\% of all models) or rocky (45\% of all models) bins. Even after reducing the uncertainties to 20\% of their current values neither of the rocky nor not rocky bins reach the 75\% threshold. We do consider LTT 1445 A c to be \textit{classifiable} at 20\% uncertainties as the sum of these two bins is greater than 75\%. Because the the largest fraction of models falls within the rocky bin, we would classify LTT 1445 A c as a \textit{leans} rocky exoplanet. 

Despite these lower uncertainties of 20\% in mass and radius for LTT 1445 A c we are unable to definitively classify LTT 1445 A c as being either rocky or not rocky because we do not know what the new values of $M_{avg}$ and $R_{avg}$ will be when measured at these lower uncertainties. As such, our model predicts that any new measurement at 20\% uncertainties in mass and radius will have sufficiently low uncertainties to produce a not rocky outcome 40\% of the time, a rocky outcome 45\% of the time, depending on the new values of $M_{avg}$ and $R_{avg}$. This is seen in Figure \ref{fig:example}, center, where at 20\% uncertainties in mass and radius, those models with average masses and radii that place them within the NRPZ typically all have NRPZ\% greater than 75\%, while those outside the NRPZ have NRPZ\% $<25\%$. 15\% of all new measurements would thus produce an inconclusive result, typically because the newly measured $M_{avg}$ and $R_{avg}$ values place them very close to, but not wholly within, the NRPZ, where even more stringent requirements on $\sigma_M$ and $\sigma_R$ are required to classify them as being inside or outside of the NRPZ. 

\section{Minimum Uncertainties to Classify Exoplanets}
\label{sec:best_uncertainties}
\begin{figure}[t!]
    \centering
    \includegraphics[width=\linewidth]{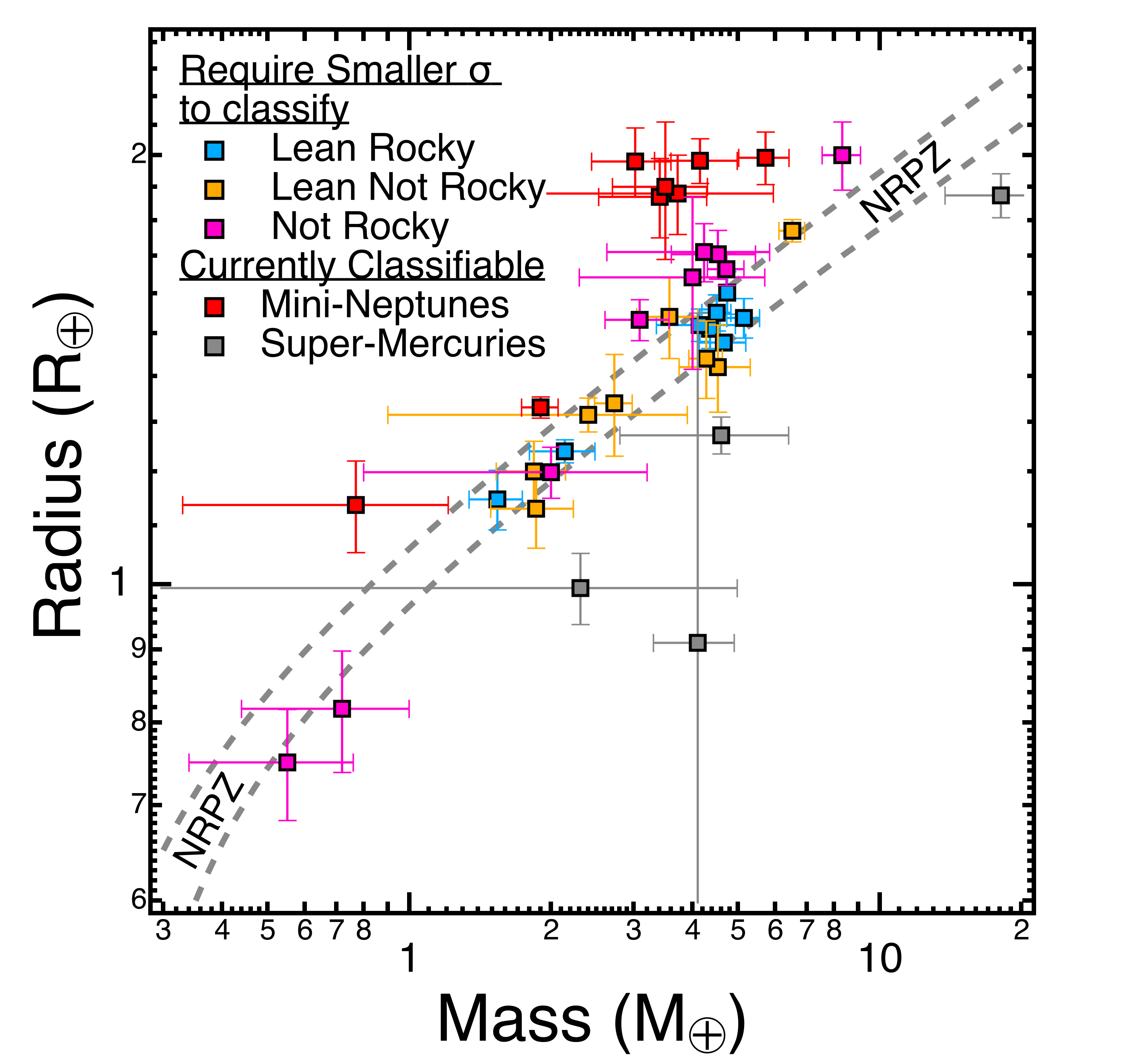}
    \caption{Current mass versus radius for the 37 planets in our sample. Twelve of these planets currently have sufficient average mass, radius and uncertainties to be classified as mini-Neptunes (red) and Super-Mercuries (gray). The remaining planets require smaller uncertainties to classify them with their exact composition being either not rocky (pink), leans rocky (blue) or leans not rocky (orange).  }
    \label{fig:MR_class}
\end{figure}
\begin{figure}
    \centering
    \includegraphics[width=\linewidth]{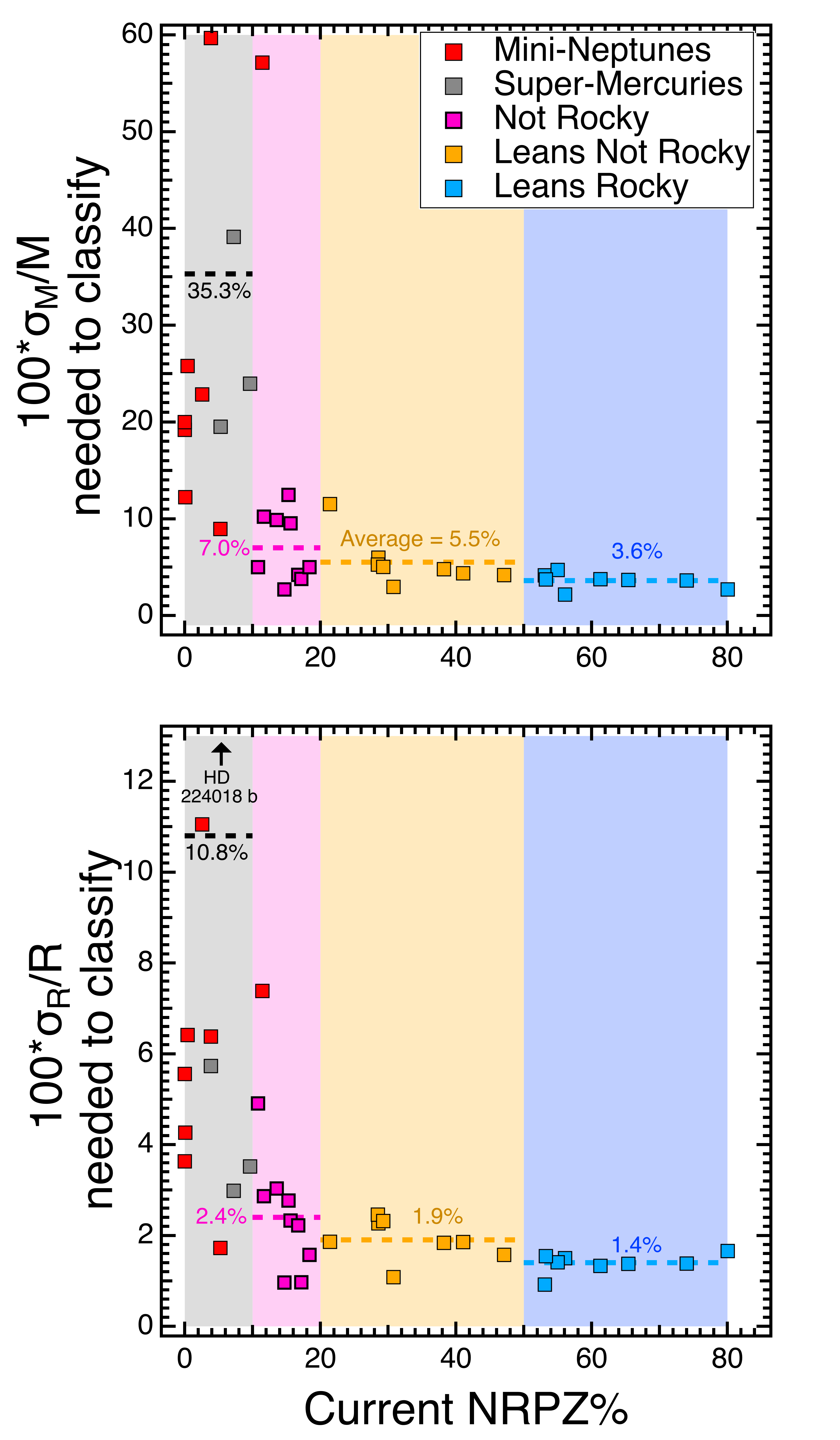}
    \caption{Average $\sigma_M$/M (top) and $\sigma_R$/R (bottom) required to classify an exoplanet as a function of the NRPZ\% based on their current uncertainties in mass and radius. Those planets with uncertainties classifiable as mini-Neptunes and super-Mercuries using current uncertainties are shown in red and gray, respectively. We also show those planets where smaller uncertainties allow us to classify them as not rocky (pink), leans not rocky (orange) and leans rocky (blue). }
    \label{fig:NRPZ_best}
\end{figure}
Following the methodology of \S\ref{sec:error_prop}, we are able to break the planets in our sample into four basic categories (Figures \ref{fig:MR_class}-\ref{fig:NRPZ_best}): those whose current uncertainties in mass and radius allow for classification, and those where improvements to $\sigma_M$ and $\sigma_R$ will allow for future classification as being either not rocky, LR or LNR. In general, we find a dependence with current NRPZ\% on the minimum uncertainties needed to classify these planets (Figure \ref{fig:NRPZ_best}). As current NRPZ\% increases, the minimum $\sigma_M$ and $\sigma_R$ relative to the current average mass and radius decreases. In general, Figure \ref{fig:NRPZ_best} shows that the NRPZ model requires less stringent uncertainties for planets that are very likely outside of the NRPZ, and is well suited to classifying not rocky planets even with large uncertainties. In order to classify a planet as rocky and within the NRPZ, however, there are tighter restrictions on the values of $M_{avg}$ and $R_{avg}$ coupled with the need for small uncertainties to increase the fraction of models that lie within the NRPZ.

\subsection{Currently Classifiable Exoplanets}
\begin{figure*}
    \centering
    \includegraphics[width=\linewidth]{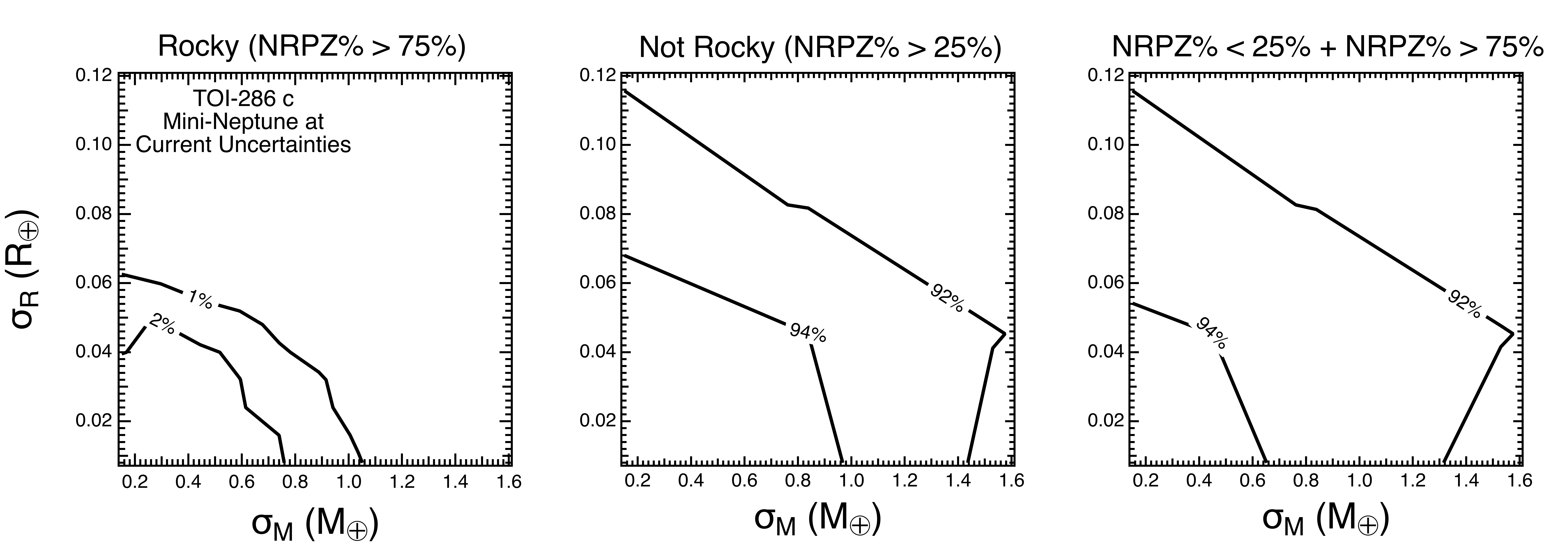}
    \caption{Contours of the percentage of models falling within the rocky (\textbf{left}) and not rocky bins (\textbf{center}) and the sum of both (\textbf{right}) bins for a given uncertainty in mass and radius for planet TOI-286 c. No best fit line is plotted as current uncertainties are sufficient to classify this planet.}
    \label{fig:TOI_MN}
\end{figure*}
Of the 37 planets in our sample, 12 have sufficient $\sigma_M$ and $\sigma_R$ to classify as definitively not rocky, 8 of which are mini-Neptunes and 4 are super-Mercuries (Table \ref{tab:MN_SM}). These planets currently have an average $\sigma_M$/M and $\sigma_R$/R of 28\% and 5.8\%, respectively. Their current average densities are sufficiently above or below the NRPZ to definitively classify them despite these high uncertainties. 

\begin{table*}
    \centering
    \caption{Currently Classifiable Exoplanets}
    \begin{tabular}{lcccc}
         & Current & Current & Current &  \\ 
        Planet & Radius (R$_\oplus$) & Mass (M$_\oplus$) & NRPZ\% & Classification \\ 
        \hline
        GJ 806 b & 1.33$\pm$0.02 (1.7\%)\tablenotemark{a} & 1.90$\pm$0.17 (8.9\%) & 5.3 & Mini-Neptune \\ 
        GJ 9827 d & 1.98$\pm$0.11 (5.6\%) & 3.02$\pm$0.58 (19.2\%) & $<0.1$ & Mini-Neptune \\ 
        HD 23472 c & 1.87$\pm$0.12 (6.4\%) & 3.41$\pm$0.88 (25.8\%) & 0.4 & Mini-Neptune \\ 
        HD 23472 f & 1.14$\pm$0.08 (7.4\%) & 0.77$\pm$0.44 (57.1\%) & 11.5 & Mini-Neptune \\ 
        HD 235088 b & 1.98$\pm$0.07 (3.6\%) & 4.15$\pm$0.83 (20.0\%) & $<0.1$ & Mini-Neptune \\ 
        TOI-286 c & 1.88$\pm$0.12 (6.4\%) & 3.72$\pm$2.22 (59.7\%) & 3.9 & Mini-Neptune \\ 
        TOI-733 b & 1.99$\pm$0.09 (4.3\%) & 5.72$\pm$0.70 (12.2\%) & 0.1 & Mini-Neptune \\ 
        TOI-1136 b & 1.90$\pm$0.21 (11.1\%) & 3.50$\pm$0.80 (22.9\%) & 2.6 & Mini-Neptune \\ 
        \hline
        HD 224018 b & 0.91$\pm$0.64 (70.3\%) & 4.10$\pm$0.80 (19.5\%) & 5.3 & Super-Mercury \\ 
        K2-233 c & 1.27$\pm$0.04 (3.0\%) & 4.60$\pm$1.80 (39.1\%) & 7.2 & Super-Mercury \\ 
        TOI-2322 b & 0.99$\pm$0.06 (5.7\%) & 2.31$\pm$2.67 (115.6\%) & 3.8 & Super-Mercury \\ 
        TOI-2322 c & 1.87$\pm$0.07 (3.5\%) & 18.10$\pm$4.34 (24.0\%) & 9.6 & Super-Mercury \\ 
    \end{tabular}
    \tablenotetext{a}{Relative uncertainty in radius or mass, $100*\sigma_X$/X}
    \label{tab:MN_SM}
\end{table*}
As an example of one of these currently classifiable planets we show results for TOI-286 c, which has a mass of 6.4$\pm$3.7 M$_\oplus$, radius 1.88$\pm$0.12 R$_\oplus$ and a current NRPZ\% of $\sim4\%$. At its current uncertainties, $\sim$90\% of all models fall within the not rocky bin (NRPZ\%$<25\%$, Figure \ref{fig:TOI_MN}), with the remaining 10\% of models being inconclusive (25\%$\leq$NRPZ\%$\leq50\%$). As uncertainties drop, the percentage of inconclusive models drops due to the number of models classifiable as rocky (NRPZ\%$\geq$75\%) increasing to $\sim1-2\%$. This is due to the large initial error bars of TOI-286 c producing $M_{avg}$ and $R_{avg}$ that fall within the NRPZ, which combined with lower uncertainties, would increase the NRPZ\% of that model to above 75\%. Regardless, the vast majority of models fall outside of the NRPZ, and based on the current $M_{avg}$ and $R_{avg}$ of TOI-286 c placing it above the NRPZ in mass-radius space, we can classify it as a mini-Neptune. 

Similar logic is followed for the remaining ten planets that fall definitively outside the NRPZ currently. Overall, these planets show the utility of the NRPZ in classifying planets that are definitively \textit{not rocky}, due to the outlier $M_{avg}$ and $R_{avg}$ values (relative to the NRPZ) compared to the rest of our sample.

\subsection{Planets Requiring Lower Uncertainties to Classify}
\begin{table*}
    \centering
    \caption{Uncertainties needed to classify exoplanets in our sample}
    \begin{tabular}{lccc|cc}
     & Current & Current & Current & Average Minimum& Average Minimum  \\
    Planet & Radius (R$_\oplus$) & Mass (M$_\oplus$) & NRPZ\% & $\sigma_R$ to Classify & $\sigma_M$ to Classify \\
    \hline
    \multicolumn{6}{c}{Confirmable Not Rocky Exoplanets}\\
    \hline
    HD 23472 b & 2.00$\pm$0.11 (5.5\%)\tablenotemark{a} & 8.32$\pm$0.78 (9.4\%) & 10.81 & 0.10 (4.9\%) & 0.42 (5.0\%) \\ 
    HD 23472 d & 0.75$\pm$0.07 (8.9\%) & 0.55$\pm$0.21 (38.2\%) & 14.702 & 0.01 (1.0\%) & 0.01 (2.7\%) \\ 
    HD 23472 e & 0.82$\pm$0.08 (9.8\%) & 0.72$\pm$0.28 (38.9\%) & 16.727 & 0.02 (2.2\%) & 0.03 (4.2\%) \\ 
    HD 136352 b & 1.66$\pm$0.04 (2.6\%) & 4.72$\pm$0.42 (8.9\%) & 17.187 & 0.02 (1.0\%) & 0.18 (3.8\%) \\ 
    HD 260655 c & 1.53$\pm$0.05 (3.3\%) & 3.09$\pm$0.48 (15.5\%) & 11.652 & 0.04 (2.9\%) & 0.32 (10.2\%) \\ 
    TOI-260 b & 1.71$\pm$0.08 (4.7\%) & 4.23$\pm$1.60 (37.8\%) & 15.283 & 0.05 (2.8\%) & 0.53 (12.5\%) \\ 
    TOI-836 b & 1.70$\pm$0.07 (3.9\%) & 4.53$\pm$0.92 (20.3\%) & 13.574 & 0.05 (3.0\%) & 0.45 (9.9\%) \\ 
    TOI-1411 b & 1.20$\pm$0.05 (4.1\%) & 2.00$\pm$1.20 (60.0\%) & 18.388 & 0.02 (1.6\%) & 0.10 (5.0\%) \\ 
    TOI-1799 b & 1.64$\pm$0.23 (13.8\%) & 4.00$\pm$1.70 (42.5\%) & 15.603 & 0.04 (2.3\%) & 0.49 (12.1\%) \\ 
    \hline
    \multicolumn{6}{c}{Confirmable Leans Not Rocky Exoplanets}\\
    \hline
    GJ 357 b & 1.20$\pm$0.06 (5.0\%) & 1.84$\pm$0.31 (16.8\%) & 47.083 & 0.02 (1.6\%) & 0.08 (4.2\%) \\ 
    GJ 9827 b & 1.44$\pm$0.09 (6.3\%) & 4.28$\pm$0.35 (8.2\%) & 41.051 & 0.03 (1.9\%) & 0.19 (4.4\%) \\ 
    GJ 9827 c & 1.13$\pm$0.07 (6.2\%) & 1.86$\pm$0.37 (19.9\%) & 29.261 & 0.03 (2.3\%) & 0.09 (5.0\%) \\ 
    HD 15337 b & 1.77$\pm$0.03 (1.8\%) & 6.52$\pm$0.41 (6.3\%) & 30.763 & 0.02 (1.1\%) & 0.19 (3.0\%) \\ 
    K2-233 b & 1.32$\pm$0.04 (2.7\%) & 2.40$\pm$1.50 (62.5\%) & 21.411 & 0.02 (1.9\%) & 0.28 (11.5\%) \\ 
    LTT 1445 A b & 1.34$\pm$0.11 (8.2\%) & 2.73$\pm$0.25 (9.2\%) & 38.213 & 0.02 (1.8\%) & 0.13 (4.8\%) \\ 
    TOI-286 b & 1.42$\pm$0.10 (7.0\%) & 4.53$\pm$0.78 (17.2\%) & 28.468 & 0.03 (2.5\%) & 0.24 (5.3\%) \\ 
    TOI-512 b & 1.54$\pm$0.10 (6.5\%) & 3.57$\pm$0.53 (14.8\%) & 28.555 & 0.03 (2.3\%) & 0.21 (6.0\%) \\ 
    \hline
    \multicolumn{6}{c}{Confirmable Leans Rocky Exoplanets}\\
    \hline
    HD 219134 b & 1.60$\pm$0.06 (3.4\%) & 4.74$\pm$0.19 (4.0\%) & 56.065 & 0.02 (1.5\%) & 0.10 (2.2\%) \\ 
    HD 219134 c & 1.51$\pm$0.05 (3.1\%) & 4.36$\pm$0.22 (5.0\%) & 80.058 & 0.03 (1.7\%) & 0.12 (2.7\%) \\ 
    HD 260655 b & 1.24$\pm$0.02 (1.9\%) & 2.14$\pm$0.34 (15.9\%) & 61.301 & 0.02 (1.3\%) & 0.08 (3.8\%) \\ 
    HIP 29442 c & 1.55$\pm$0.05 (2.9\%) & 4.50$\pm$0.32 (7.1\%) & 73.998 & 0.02 (1.4\%) & 0.16 (3.6\%) \\ 
    HIP 29442 d & 1.54$\pm$0.05 (3.2\%) & 5.14$\pm$0.41 (8.0\%) & 65.409 & 0.02 (1.4\%) & 0.19 (3.7\%) \\ 
    Kepler-65 d & 1.52$\pm$0.04 (2.6\%) & 4.14$\pm$0.79 (19.1\%) & 54.995 & 0.02 (1.4\%) & 0.20 (4.7\%) \\ 
    Kepler-93 b & 1.48$\pm$0.02 (1.3\%) & 4.66$\pm$0.53 (11.4\%) & 53.09 & 0.01 (0.9\%) & 0.19 (4.2\%) \\ 
    LTT 1445 A c & 1.15$\pm$0.11 (9.6\%) & 1.54$\pm$0.25 (16.2\%) & 53.233 & 0.02 (1.5\%) & 0.06 (3.7\%) \\ 

    \end{tabular}
    \tablenotetext{a}{Relative uncertainty in radius or mass, $100*\sigma_X$/X}
    \label{tab:plan_best}

\end{table*}
The remaining planets in our sample all require smaller uncertainties to definitively classify them as either confirmable not rocky (CNR), LR or LNR planets. These planets typically have $M_{avg}$ and $R_{avg}$ values that are near or within the NRPZ (Figure \ref{fig:MR_class}) compared to those that we can currently classify as not rocky described above. Their high current uncertainties, however, lower their current NRPZ\% such that they are currently inconclusive. Through our analysis, we found smaller uncertainties yielded an increase in the fraction of models that fell into either the rocky (NRPZ\% $>$ 75\%) and not rocky (NRPZ\% $<$ 25\%) bins. Of these remaining planets, 9 cross the 75\% threshold for definitive classification as being not rocky, with 8 each falling into the lean not rocky and lean rocky categories. No planet in our sample crosses the 75\% threshold for being confirmed as rocky exoplanets when lower uncertainties are applied. 

For each of these planets, the contours of the resulting percentage of all models within the rocky (NRPZ\%$>75\%$), not rocky (NRPZ\% $<$ 25\%) bins and the sum of the two as a function of lower mass and radius uncertainties are shown in Supplemental Figures \ref{fig:app_CNR}, \ref{fig:app_NR} and \ref{fig:app_R} for planets we classify as confirmable not rocky (CNR), LNR and LR, respectively. To select the minimum $\sigma_M$ and $\sigma_R$ needed to classify these planets, we fit the 75\% contour of either the not rocky or sum of the rocky and not rocky bins and selected the midpoint of this fit. The minimum uncertainties reported in Table \ref{tab:plan_best} are these midpoints and are meant only as a guide as any uncertainty on the 75\% contours will produce a result that allows the planet to be classified. 

\subsubsection{Confirmable Not Rocky Exoplanets}
\begin{figure*}
    \centering
    \includegraphics[width=\linewidth]{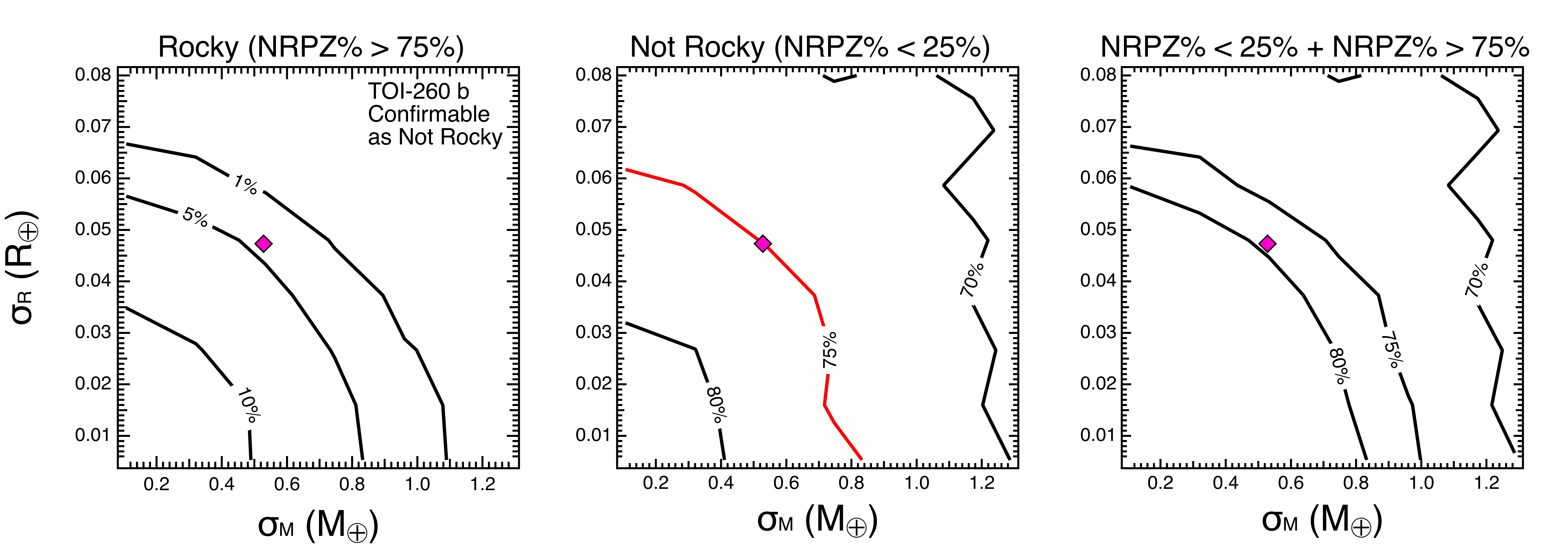}
    \caption{Contours of the percentage of models falling within the rocky (\textbf{left}) and not rocky bins (\textbf{center}) and the sum of both (\textbf{right}) bins for a given uncertainty in mass and radius for planet TOI-260 b. Our best-fit uncertainty to classify this planet as not rocky is shown as a pink diamond.  }
    \label{fig:TOI_CNR}
\end{figure*}
We are able to classify 8 planets as confirmable not rocky (CNR) when lower uncertainties are applied (Table \ref{tab:plan_best}). As modeled uncertainties decrease, each of these planets reach 75\% of all models being within the not rocky bin (NRPZ\%$<25\%$). Each of these planets each have current NRPZ\% values between $\sim10-20\%$ (Figure \ref{fig:NRPZ_best}). We find average best fit $\sigma_M/M$ and $\sigma_R/R$ of 7.3\% and 2.4\%, respectively (Figure \ref{fig:NRPZ_best}). This represents a $\sim67\%$ and $\sim51\%$ average reduction in the current mass and radius uncertainties, respectively. We provide contours for each of the CNR planets in Supplemental Figure \ref{fig:app_CNR}. 

One planet, TOI-1411 b has a current average mass and radius that places it within the NRPZ, however, its relative uncertainties in mass and radius of 60\% and 4.1\%, respectively, lower its current NRPZ\% to just $\sim18\%$. Due to these large current relative uncertainties for TOI-1411 b, particularly in mass, our analysis selects a large amount of $M_{avg}$ and $R_{avg}$ values that lie outside of the NRPZ, thus increasing the proportion of models that produce outcomes within the NRPZ\% $<$25\% bin. Thus even as uncertainties drop, our model shows that 75\% of models still fall outside of the NRPZ with sufficient uncertainties to classify it as CNR. 

For example, we show contours of the percentage of all models within the not rocky (NRPZ\%$ < 25\%$), rocky (NRPZ\% $> 75\%$) and the sum of the two bins of the CNR exoplanet TOI-260 b in Figure \ref{fig:TOI_CNR}. For this planet, we estimate that 75\% of all models have NRPZ\% $<$ 25\% when $\sigma_M$ is below 0.8 M$_\oplus$ and $\sigma_R$ is below 0.06 R$_\oplus$. We select a best-fit uncertainties to classify TOI-260 b of $\sigma_{M}\approx0.53$ M$_\oplus$ and $\sigma_{R}\approx0.05$ R$_\oplus$. These lower uncertainties represent a decrease of approximately 67\% and 41\% from the current uncertainties in mass and radius, respectively. If a new measurement is achieved at these uncertainties, we estimate that 75\% of new measurements would produce $M_{avg}$ and $R_{avg}$ that fall outside of the NRPZ and with sufficient uncertainty to give it an NRPZ\%$<25\%$, thus classifying it as a not rocky exoplanet. Additionally, we estimate $\sim4\%$ of new measurements would produce $M_{avg}$ and $R_{avg}$ values that fall within the NRPZ, with sufficient uncertainty to give it an NRPZ\%$>75\%$, thus classifying it as a rocky exoplanet. Finally, we estimate $\sim21\%$ of new measurements at these lower uncertainties would produce a planet with $M_{avg}$ and $R_{avg}$ that fall very near the NRPZ. In this latter case, the resulting NRPZ\% would be between $25\%-75\%$, thus producing an inconclusive result, despite the uncertainties of the new measurement being small. If an inconclusive classification is the outcome, even smaller uncertainties than those we report would be required to classify TOI-260 b as either rocky or not rocky. 

\subsubsection{Leans Not Rocky Exoplanets}
We classify planets as LNR when neither the rocky nor not rocky bin reaches the 75\% threshold and the largest fraction of models falls within the NRPZ \% $<25\%$ bin. These planets typically have current $M_{avg}$ and $R_{avg}$ values that fall outside of the NRPZ (Figure \ref{fig:MR_class}, orange points). Two planets, however, have $M_{avg}$ and $R_{avg}$ that fall within the NRPZ: GJ 357 b and TOI-1411 b. Due to the large current relative uncertainties for these planets, particularly in mass, our analysis selects a large amount of $M_{avg}$ and $R_{avg}$ values that lie outside the NRPZ, thus increasing the proportion of models that produce results with NRPZ \% $<$25\%. In total, we find 8 planets where smaller uncertainties will allow us to classify these planets as LNR. On average, we estimate these leans not rocky exoplanets require average $\sigma_M/M$ and $\sigma_R/R$ of 5.5\% and 1.9\%, respectively (Figure \ref{fig:NRPZ_best}). These uncertainties represent an average $\sim60\%$ reduction in the current mass and radius uncertainties. We provide contours for each of the leans rocky exoplanets in Supplemental Figure \ref{fig:app_NR}.

\begin{figure*}
    \centering
    \includegraphics[width=\linewidth]{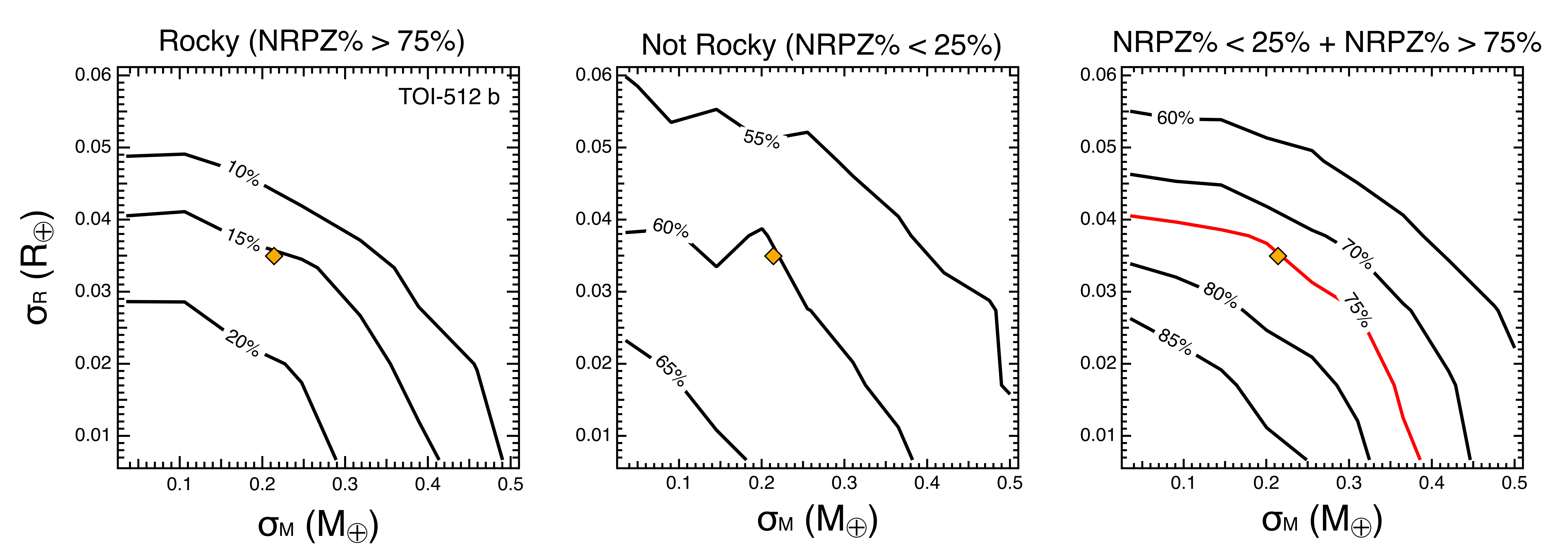}
    \caption{Contours of the percentage of models falling within the rocky (\textbf{left}) and not rocky bins (\textbf{center}) and the sum of both (\textbf{right}) bins for a given uncertainty in mass and radius for planet TOI-512 b. Our best-fit uncertainty to classify this planet as not rocky is shown as an orange diamond.}
    \label{fig:TOI_NR}
\end{figure*}
For example, we show contours percentage of all model within the not rocky (NRPZ\%$ < 25\%$), rocky (NRPZ\% $> 75\%$) and the sum of the two bins of the LNR exoplanet TOI-512 b in Figure \ref{fig:TOI_NR}. Neither the rocky or not rocky bins reach our threshold of 75\% of all models being within either as uncertainties decrease. The sum of the two bins does, however, reach this 75\% threshold when mass uncertainties are below $\sim$0.4 M$_\oplus$ and radius uncertainties below $\sim0.04$ R$_\oplus$. We select best-fit uncertainties for this planet to classify it as LNR at $\sigma_M\approx0.21$ M$_\oplus$ and $\sigma_R\approx0.04$ R$_\oplus$. Measurements at these lower uncertainties would represent a 59\% and 65\% reduction in the current uncertainties in mass and radius, respectively. A new measurement at these uncertainties would, we predict, have an $\sim15\%$ likelihood of producing a rocky planet with NRPZ\%$>75\%$, a $\sim60\%$ likelihood of producing a not rocky exoplanet with NRPZ\%$<25\%$ and 25\% likelihood of producing an inconclusive result with $25\%<\rm{NRPZ\%}<75\%$. If a new measurement does produce an inconclusive outcome, even smaller uncertainties would be necessary to classify TOI-512 b as either rocky or not rocky.

\subsubsection{Leans Rocky Exoplanets}
We classify planets as LR when neither the rocky or not rocky bin reach the 75\% threshold and the largest fraction of models falls within the NRPZ\% $>75\%$ bin. These planets all have current $M_{avg}$ and $R_{avg}$ values that are within the NRPZ (Figure \ref{fig:MR_class}, blue points). In total, we find 8 planets where smaller uncertainties will allow us to classify these planets as LR exoplanets. These planets all have current NRPZ\% between 50 and 80\%. We report the best-fit uncertainties needed to classify these planets as LR in Table \ref{tab:plan_best}. We estimate that these planets require an average best-fit $\sigma_M/M$ and $\sigma_R/R$ of 3.6\% and 1.4\%, respectively (Figure \ref{fig:NRPZ_best}). Our best-fit uncertainties represent an average of $\sim60\%$ and $\sim$48\% reduction in current mass and radius uncertainties, respectively, across all planets in our LR subsample. We provide contours of the percentage of models that fall within each bin for each of the planets we can classify as LR in Supplemental Figure \ref{fig:app_R}.  

\begin{figure*}
    \centering
    \includegraphics[width=\linewidth]{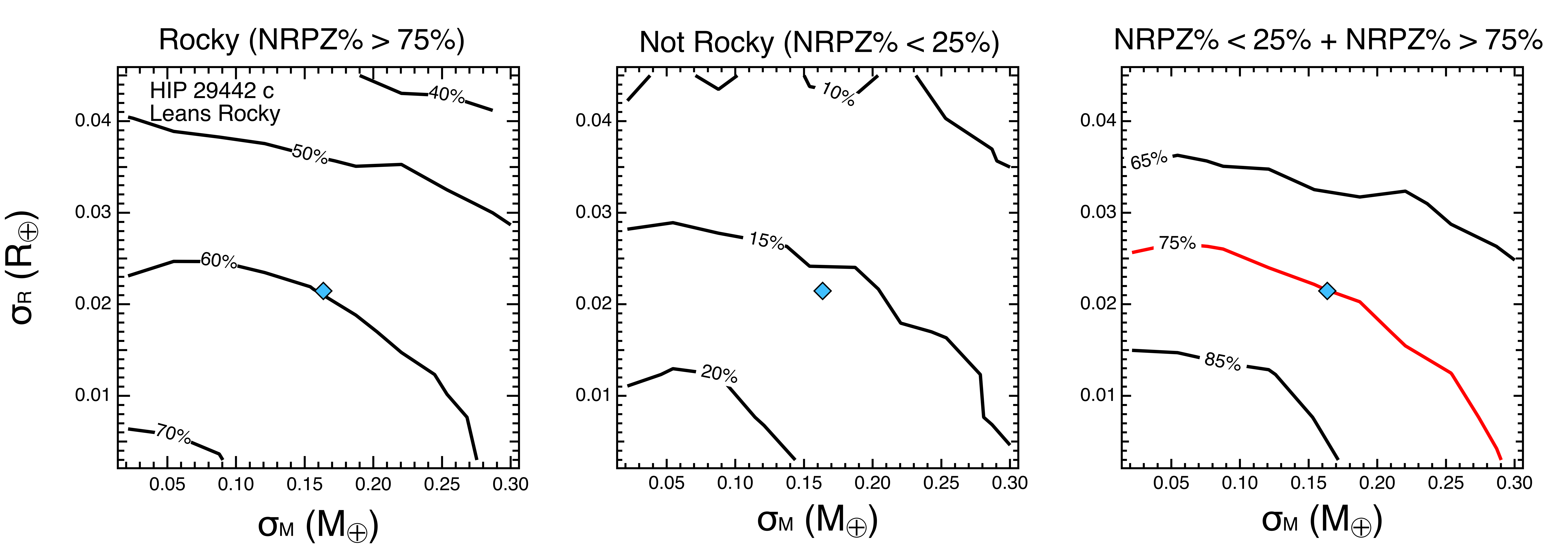}
    \caption{Contours of the percentage of models falling within the rocky (\textbf{left} and not rocky bins (\textbf{center}) and the sum of both (\textbf{right}) bins for a given uncertainty in mass and radius for planet HIP 29442 c. Our best-fit uncertainty to classify this planet as not rocky is shown as an blue diamond.}
    \label{fig:HIP_R}
\end{figure*}

For example, we show contours percentage of all model within the not rocky (NRPZ\%$ < 25\%$), rocky (NRPZ\% $> 75\%$) and the sum of the two bins of the leans rocky exoplanet HIP 29442 c in Figure \ref{fig:HIP_R}. Neither the rocky or not rocky bins reach our threshold of 75\% of all models being within either the rocky or not rocky bin as uncertainties decrease. The sum of the two bins does reach this 75\% threshold, however, when mass uncertainties are below $\sim$0.3 M$_\oplus$ and radius uncertainties below $\sim0.025$ R$_\oplus$. We select best-fit uncertainties for this planet to classify it as a LR at $\sigma_M\approx0.16$ M$_\oplus$ and $\sigma_R\approx0.02$ R$_\oplus$. Measurements at these lower uncertainties would represent a 48\% and 52\% reduction in the current uncertainties in mass and radius, respectively. A new measurement at these uncertainties would, we predict, have an $\sim60\%$ likelihood of producing a rocky planet with NRPZ\%$>75\%$, a $\sim15\%$ likelihood of producing a not rocky exoplanet with NRPZ\%$<25\%$ and 25\% likelihood of producing an inconclusive result with $25\%<\rm{NRPZ\%}<75\%$. If a new measurement does produce an inconclusive outcome, even smaller uncertainties would be necessary to classify HIP 29442 c as either rocky or not rocky. 

\section{Observational Requirements for Lowering Mass-Radius Uncertainties}
\label{sec:feasibility}
With the results of \S\ref{sec:best_uncertainties} in hand, we can estimate the number of new measurements required to reach the mass and radius precisions needed for NRPZ classification. We emphasize that these estimates refer to statistically independent measurements, not to the number of planetary orbits. Multiple RV measurements may be obtained during a single orbit, and a single transit generally contains many photometric measurements. Converting the measurement counts below into telescope time or calendar duration requires an explicit observing cadence, exposure time, phase-coverage strategy, and treatment of correlated noise.

\subsection{Mass Uncertainties}

Following the Fisher-information treatment of \citet{cloutier2018b}, and generalizing it to arbitrary orbital-phase sampling and a fitted systemic velocity offset, we write the velocity at time $t_i$ for an RV orbit whose ephemeris and shape are known as
\begin{equation}
    v_i = \gamma + K g_i, \qquad
    g_i = \cos(\nu_i+\omega)+e\cos\omega,
    \label{eq:rv_model}
\end{equation}
where $\gamma$ is a constant velocity offset, $K$ is the RV semi-amplitude, $\nu_i$ is the true anomaly, $e$ is the eccentricity, and $\omega$ is the argument of periastron. For $N_{\rm RV}$ measurements with equal, independent per-measurement uncertainty $\sigma_{\rm RV,1}$, the formal uncertainty in $K$ after fitting $\gamma$ is
\begin{equation}
    \sigma_K^2 \simeq
    \frac{\sigma_{\rm RV,1}^2}
    {\sum_{i=1}^{N_{\rm RV}}(g_i-\overline{g})^2}
    = \frac{\sigma_{\rm RV,1}^2}
    {N_{\rm RV}\eta_{\rm RV}},
    \label{eq:sigma_K}
\end{equation}
where
\begin{equation}
    \eta_{\rm RV} \equiv
    \frac{1}{N_{\rm RV}}
    \sum_{i=1}^{N_{\rm RV}}(g_i-\overline{g})^2
    \label{eq:eta_RV}
\end{equation}
is the phase-leverage factor. For a circular orbit sampled uniformly in phase, $\eta_{\rm RV}\simeq1/2$ and $\sigma_K\simeq\sigma_{\rm RV,1}\sqrt{2/N_{\rm RV}}$. Balanced measurements near the positive and negative velocity quadratures can approach $\eta_{\rm RV}\simeq1$, whereas clustered or otherwise unfavorable phase coverage yields smaller $\eta_{\rm RV}$. The importance of temporal sampling and observing strategy in RV sensitivity and parameter recovery has previously been discussed extensively in the literature \citep{cumming2004,kane2007a,kane2007b,gupta2021,li2025g}.

For $M_p\ll M_\star$, the RV-derived planet mass scales as $M_p\propto K M_\star^{2/3}P^{1/3}(1-e^2)^{1/2}/\sin i$. Neglecting covariances, the target fractional mass uncertainty $f_M\equiv\sigma_M/M_p$ can therefore be approximated as
\begin{equation}
    f_M^2 \simeq
    \left(\frac{\sigma_K}{K}\right)^2 + f_{M,0}^2,
    \label{eq:sigma_M_full}
\end{equation}
where the uncertainty floor contributed by the remaining stellar and orbital parameters is
\begin{equation}
\begin{array}{rcl}
    f_{M,0}^2 &\simeq&
    \displaystyle\frac{4}{9}\left(\frac{\sigma_{M_\star}}{M_\star}\right)^2
    + \displaystyle\frac{1}{9}\left(\frac{\sigma_P}{P}\right)^2 \\
    &&+ \displaystyle\left(\frac{e\sigma_e}{1-e^2}\right)^2
    + \left(\cot i\,\sigma_i\right)^2.
\end{array}
    \label{eq:mass_floor}
\end{equation}
This term includes the uncertainties in stellar mass, orbital period, eccentricity, and inclination, but excludes the uncertainty in the RV semi-amplitude $K$. Here $\sigma_i$ is expressed in radians. Combining Equations~\ref{eq:sigma_K} and \ref{eq:sigma_M_full}, the idealized number of independent RV measurements required is
\begin{equation}
    N_M \equiv N_{\rm RV} \simeq
    \frac{\sigma_{\rm RV,1}^2}
    {\eta_{\rm RV}K^2\left(f_M^2-f_{M,0}^2\right)}.
    \label{eq:N_M}
\end{equation}
This expression is finite only when $f_M>f_{M,0}$. Period and inclination uncertainties are generally small for transiting planets, but stellar-mass and eccentricity uncertainties can impose non-negligible floors. In addition, stellar activity, granulation, instrumental systematics, telluric contamination, and multi-planet covariances can prevent the uncertainty from following the ideal $N_{\rm RV}^{-1/2}$ scaling \citep{fischer2016,luhn2023}.

Table~\ref{tab:mass_err} adopts a per-measurement precision of $\sigma_{\rm RV,1}=0.5$~m~s$^{-1}$ as an optimistic benchmark, comparable to the median internal precisions of 0.35 and 0.48~m~s$^{-1}$ achieved with MAROON-X for GJ~12 \citep{brady2025}. We neglect $f_{M,0}$, so the tabulated values remain optimistic lower limits. Because Equation~\ref{eq:N_M} gives $N_{\rm RV}\propto\sigma_{\rm RV,1}^{2}$, the tabulated counts can be rescaled straightforwardly for other assumed single-measurement precisions. We report both balanced quadrature sampling ($\eta_{\rm RV}=1$) and uniform phase sampling ($\eta_{\rm RV}=1/2$). Under uniform phase sampling, the planets in our sample require approximately 10--8100 RV measurements for the confirmable not-rocky planets, 20--180 measurements for the LNR planets, and 80--240 measurements for the LR planets. Optimized quadrature sampling can reduce these counts by approximately a factor of two, but real programs must also obtain sufficient phase and temporal coverage to distinguish planetary signals from activity and other planets. Consequently, multiplying $N_{\rm RV}$ by the orbital period does not provide a valid observing duration.

\begin{table*}
    \centering
    \caption{Idealized RV measurement requirements for classification with improved mass uncertainties}
    \begin{tabular}{lcccccc}
         & RV & Orbital & Current & Required & \multicolumn{2}{c}{$N_{\rm RV}$ measurements needed\tablenotemark{a}} \\
         \cline{6-7}
         Planet & $K$ (m s$^{-1}$) & Period (days) & Mass (M$_\oplus$) & $\sigma_M$ (M$
_\oplus$) & $\eta_{\rm RV}=1$ & $\eta_{\rm RV}=1/2$ \\
         \hline
        \multicolumn{7}{c}{Confirmable Not Rocky Exoplanets}\\
        \hline
        HD 23472 b & 2.68 & 17.7 & 8.32$\pm$0.78 & 0.42 & 14 & 28 \\
        TOI-836 b & 2.38 & 3.8 & 4.53$\pm$0.92 & 0.45 & 5 & 10 \\
        HD 260655 c & 1.92 & 5.7 & 3.09$\pm$0.48 & 0.32 & 6 & 12 \\
        TOI-260 b & 1.57 & 13.5 & 4.23$\pm$1.60 & 0.53 & 7 & 14 \\
        TOI-1411 b & 1.55 & 1.5 & 2.00$\pm$1.20 & 0.10 & 41 & 82 \\
        HD 136352 b & 1.46 & 11.6 & 4.72$\pm$0.42 & 0.18 & 82 & 164 \\
        TOI-1799 b & 1.40 & 7.1 & 4.00$\pm$1.70 & 0.49 & 9 & 18 \\
        HD 23472 e & 0.30 & 7.9 & 0.72$\pm$0.28 & 0.03 & 1555 & 3110 \\
        HD 23472 d & 0.29 & 4.0 & 0.55$\pm$0.21 & 0.01 & 4053 & 8106 \\
        \hline
        \multicolumn{7}{c}{Confirmable Leans Not Rocky Exoplanets}\\
        \hline
        GJ 9827 b & 3.53 & 1.2 & 4.28$\pm$0.35 & 0.19 & 10 & 20 \\
        HD 15337 b & 2.82 & 4.8 & 6.52$\pm$0.41 & 0.19 & 35 & 70 \\
        LTT 1445 A b & 2.47 & 5.4 & 2.73$\pm$0.25 & 0.13 & 18 & 36 \\
        GJ 357 b & 2.13 & 3.9 & 1.84$\pm$0.31 & 0.08 & 31 & 62 \\
        TOI-286 b & 1.98 & 4.5 & 4.53$\pm$0.78 & 0.24 & 23 & 46 \\
        TOI-512 b & 1.54 & 7.2 & 3.57$\pm$0.53 & 0.21 & 29 & 58 \\
        K2-233 b & 1.31 & 2.5 & 2.40$\pm$1.50 & 0.28 & 11 & 22 \\
        GJ 9827 c & 1.06 & 3.6 & 1.86$\pm$0.37 & 0.09 & 88 & 176 \\
        \hline
        \multicolumn{7}{c}{Confirmable Leans Rocky Exoplanets}\\
        \hline
        HD 219134 b & 2.38 & 3.1 & 4.74$\pm$0.19 & 0.10 & 92 & 184 \\
        HIP 29442 c & 2.04 & 3.5 & 4.50$\pm$0.32 & 0.16 & 45 & 90 \\
        HIP 29442 d & 1.91 & 6.4 & 5.14$\pm$0.41 & 0.19 & 50 & 100 \\
        Kepler-93 b & 1.89 & 4.7 & 4.66$\pm$0.53 & 0.19 & 40 & 80 \\
        HD 219134 c & 1.70 & 6.8 & 4.36$\pm$0.22 & 0.12 & 120 & 240 \\
        HD 260655 b & 1.69 & 2.8 & 2.14$\pm$0.34 & 0.08 & 62 & 124 \\
        LTT 1445 A c & 1.67 & 3.1 & 1.54$\pm$0.20 & 0.06 & 64 & 128 \\
        Kepler-65 d & 1.17 & 8.1 & 4.14$\pm$0.79 & 0.20 & 83 & 166 \\
    \end{tabular}
    \tablenotetext{a}{Assumes independent 0.5~m~s$^{-1}$ measurements and $f_{M,0}=0$. The $\eta_{\rm RV}=1$ column represents balanced observations near both velocity quadratures; $\eta_{\rm RV}=1/2$ represents uniform phase coverage. Correlated noise and fitted nuisance parameters generally increase the required counts.}
    \label{tab:mass_err}
\end{table*}
\subsection{Radius Uncertainties}

In addition to precise masses, robust NRPZ classifications require precise planetary radii. For transiting systems, the primary observable is the transit depth, $\delta\simeq(R_p/R_\star)^2$, such that $R_p=R_\star\sqrt{\delta}$. Assuming that the stellar-radius and transit-depth uncertainties are independent, the fractional planetary-radius uncertainty is
\begin{equation}
    \left(\frac{\sigma_{R_p}}{R_p}\right)^2 \simeq
    \left(\frac{\sigma_{R_\star}}{R_\star}\right)^2 +
    \frac{1}{4}\left(\frac{\sigma_\delta}{\delta}\right)^2.
    \label{eq:sigma_R}
\end{equation}
Analytic treatments of transit observables, uncertainties, and covariances are provided by \citet{seager2003a} and \citet{carter2008}. We now define $\sigma_{f,1}$ to be the relative-flux uncertainty of one photometric measurement. For a box-shaped transit with $N_{\rm in}$ in-transit measurements and $N_{\rm out}$ out-of-transit measurements of equal precision, the depth uncertainty is approximately
\begin{equation}
    \sigma_\delta^2 \simeq \sigma_{f,1}^2
    \left(\frac{1}{N_{\rm in}}+\frac{1}{N_{\rm out}}\right).
    \label{eq:sigma_delta_points}
\end{equation}
If $N_R$ comparable transits are observed, with $n_{\rm in}$ and $n_{\rm out}$ measurements per transit, then
\begin{equation}
    \sigma_\delta =
    \frac{\sigma_{f,1}}
    {\sqrt{N_R n_{\rm eff,tr}}}, \qquad
    n_{\rm eff,tr}\equiv
    \left(\frac{1}{n_{\rm in}}+\frac{1}{n_{\rm out}}\right)^{-1}.
    \label{eq:neff_transit}
\end{equation}
Thus the photometric contribution improves as the inverse square root of both the number of transits and the effective number of samples per transit. If the out-of-transit baseline is known much more precisely than the in-transit flux, $n_{\rm eff,tr}\simeq n_{\rm in}$. If equal numbers of points are obtained in and out of transit, $n_{\rm eff,tr}=n_{\rm in}/2$.

For a target fractional radius precision $f_R\equiv\sigma_{R_p}/R_p$, Equations~\ref{eq:sigma_R} and \ref{eq:neff_transit} give
\begin{equation}
    N_R \simeq
    \frac{\sigma_{f,1}^2}
    {4\delta^2 n_{\rm eff,tr}
    \left[f_R^2-\left(\sigma_{R_\star}/R_\star\right)^2\right]}.
    \label{eq:R_N}
\end{equation}
Equivalently, the required total number of effective photometric samples is
\begin{equation}
    N_{\rm eff,R}\equiv N_R n_{\rm eff,tr}
    \simeq
    \frac{\sigma_{f,1}^2}
    {4\delta^2
    \left[f_R^2-\left(\sigma_{R_\star}/R_\star\right)^2\right]}.
    \label{eq:R_Neff}
\end{equation}
Both expressions are finite only when $f_R>\sigma_{R_\star}/R_\star$; otherwise improved stellar characterization is required before additional photometry can reach the target planetary-radius precision.

Table~\ref{tab:rad_err} adopts a per-measurement precision of $\sigma_{f,1}=100$~ppm and reports $N_{\rm eff,R}$, not the number of transits. Because Equation~\ref{eq:R_Neff} depends on the product $N_R n_{\rm eff,tr}$, a transit count cannot be assigned without specifying the cadence and baseline coverage of each event. Fourteen of the 25 planets currently have $\sigma_{R_\star}/R_\star$ values too large to achieve the required $f_R$. For these planets, marked in Table~\ref{tab:rad_err}, we retain the optimistic assumption $\sigma_{R_\star}/R_\star=0.95f_R$ to illustrate the photometric requirement after improved stellar characterization. The required $N_{\rm eff,R}$ values span approximately 3--31,000 for the confirmable not-rocky planets, 18--800 for the LNR planets, and 40--13,600 for the LR planets. These are effective cadence-level samples rather than orbital cycles. As an illustration, a transit observed with 50 independent in-transit measurements and 50 independent out-of-transit measurements has $n_{\rm eff,tr}=25$ and a nominal single-transit depth uncertainty of 20~ppm for $\sigma_{f,1}=100$~ppm. Under that specific cadence, the corresponding transit counts would range from 1--1257, 1--32, and 2--544 for the three groups, respectively. Other cadences can be evaluated directly using Equation~\ref{eq:R_N}. For example, Kepler-65~d has an orbital period of 8.1~days, so one 27 day TESS sector provides only three or four transit opportunities \citep{ricker2015}. Under the illustrative cadence adopted in Table~\ref{tab:rad_err}, the required 544 transits would span at least $\sim12$~years of continuous coverage, before accounting for sector visibility or data gaps.

The white-noise scaling in Equations~\ref{eq:sigma_delta_points}--\ref{eq:R_Neff} is optimistic. Limb-darkening uncertainty, finite integration time, impact-parameter covariance, stellar variability, starspot crossings, atmospheric trends, and instrumental systematics can reduce the effective number of independent measurements. Time-correlated noise in particular does not average down as $N^{-1/2}$ and can dominate high-precision transit analyses \citep{pont2006c,carter2008}. As with the RV calculation, the values in Table~\ref{tab:rad_err} should therefore be interpreted as lower limits rather than guaranteed observing requirements.

\begin{table*}
    \centering
    \caption{Idealized photometric measurement requirements for classification with improved radius uncertainties}
    \begin{tabular}{lcccccc}
        & Transit & Stellar & Current Planet & Required & Required & Illustrative \\
        Planet & Depth $\delta$ (\%) & Radius (R$_\odot$) & Radius (R$_\oplus$) & $\sigma_R$ (R$_\oplus$) & $N_{\rm eff,R}$\tablenotemark{a} & $N_R$\tablenotemark{b} \\
        \hline
        \multicolumn{7}{c}{Confirmable Not Rocky Exoplanets}\\
        \hline
        HD 260655 c & 0.102 & 0.44$\pm$0.003 & 1.53$\pm$0.051 & 0.04 & 3 & 1 \\
        TOI-260 b & 0.066 & 0.61$\pm$0.014 & 1.71$\pm$0.080 & 0.05 & 24 & 1 \\
        HD 23472 b & 0.066 & 0.71$\pm$0.020 & 2.00$\pm$0.11 & 0.10 & 4 & 1 \\
        TOI-836 b & 0.055 & 0.67$\pm$0.010 & 1.70$\pm$0.067 & 0.05 & 12 & 1 \\
        TOI-1411 b\tablenotemark{c} & 0.032 & 0.61$\pm$0.019 & 1.20$\pm$0.049 & 0.02 & 1017 & 41 \\
        TOI-1799 b\tablenotemark{c} & 0.025 & 0.96$\pm$0.046 & 1.64$\pm$0.227 & 0.04 & 778 & 32 \\
        HD 136352 b\tablenotemark{c} & 0.021 & 1.06$\pm$0.019 & 1.66$\pm$0.043 & 0.02 & 6351 & 255 \\
        HD 23472 e\tablenotemark{c} & 0.011 & 0.71$\pm$0.020 & 0.82$\pm$0.08 & 0.02 & 4199 & 168 \\
        HD 23472 d\tablenotemark{c} & 0.009 & 0.71$\pm$0.020 & 0.75$\pm$0.067 & 0.01 & 31412 & 1257 \\
        \hline
        \multicolumn{7}{c}{Leans Not Rocky Exoplanets}\\
        \hline
        LTT 1445 A b\tablenotemark{c} & 0.205 & 0.27$\pm$0.019 & 1.34$\pm$0.11 & 0.02 & 18 & 1 \\
        GJ 357 b\tablenotemark{c} & 0.106 & 0.34$\pm$0.015 & 1.20$\pm$0.06 & 0.02 & 92 & 4 \\
        GJ 9827 b\tablenotemark{c} & 0.052 & 0.58$\pm$0.030 & 1.44$\pm$0.09 & 0.03 & 279 & 12 \\
        HD 15337 b & 0.036 & 0.86$\pm$0.008 & 1.77$\pm$0.032 & 0.02 & 653 & 27 \\
        GJ 9827 c\tablenotemark{c} & 0.032 & 0.58$\pm$0.030 & 1.13$\pm$0.07 & 0.03 & 470 & 19 \\
        K2-233 b & 0.029 & 0.71$\pm$0.010 & 1.32$\pm$0.036 & 0.02 & 204 & 9 \\
        TOI-286 b\tablenotemark{c} & 0.028 & 0.78$\pm$0.036 & 1.42$\pm$0.10 & 0.03 & 551 & 23 \\
        TOI-512 b\tablenotemark{c} & 0.025 & 0.89$\pm$0.030 & 1.54$\pm$0.10 & 0.03 & 791 & 32 \\
        \hline
        \multicolumn{7}{c}{Leans Rocky Exoplanets}\\
        \hline
        LTT 1445 A c\tablenotemark{c} & 0.157 & 0.27$\pm$0.011 & 1.15$\pm$0.055 & 0.02 & 44 & 2 \\
        HD 260655 b & 0.067 & 0.44$\pm$0.003 & 1.24$\pm$0.023 & 0.02 & 43 & 2 \\
        HD 219134 b & 0.036 & 0.78$\pm$0.005 & 1.60$\pm$0.055 & 0.02 & 107 & 5 \\
        HD 219134 c & 0.032 & 0.78$\pm$0.005 & 1.51$\pm$0.047 & 0.03 & 107 & 5 \\
        Kepler-93 b\tablenotemark{c} & 0.022 & 0.92$\pm$0.011 & 1.48$\pm$0.019 & 0.01 & 6434 & 258 \\
        HIP 29442 c & 0.021 & 0.98$\pm$0.007 & 1.55$\pm$0.045 & 0.02 & 404 & 17 \\
        HIP 29442 d & 0.021 & 0.98$\pm$0.007 & 1.54$\pm$0.049 & 0.02 & 424 & 17 \\
        Kepler-65 d\tablenotemark{c} & 0.010 & 1.41$\pm$0.030 & 1.52$\pm$0.040 & 0.02 & 13592 & 544 \\
    \end{tabular}
    \tablenotetext{a}{Required total effective photometric samples from Equation~\ref{eq:R_Neff}, assuming independent 100~ppm measurements. This is not the number of transits.}
    \tablenotetext{b}{Illustrative number of transits for $n_{\rm in}=n_{\rm out}=50$, so that $n_{\rm eff,tr}=25$.}
    \tablenotetext{c}{The current stellar-radius uncertainty is too large to achieve the desired $\sigma_{R_p}$. The calculation assumes improved stellar characterization with $\sigma_{R_\star}/R_\star=0.95f_R$.}
    \label{tab:rad_err}
\end{table*}

\section{Discussion}
\label{sec:discussion}
\subsection{Efficacy of Classifying Planets via Mass-Radius models}

The observational requirements derived in Section~\ref{sec:feasibility} show that refining mass and radius can require a substantial investment for some targets. The values in Tables~\ref{tab:mass_err} and \ref{tab:rad_err} are numbers of statistically independent RV measurements and effective photometric samples, rather than telescope days or planetary orbits. The calendar time depends on cadence, exposure time, phase coverage, visibility, and correlated noise. Both current EPRV instruments such as ESPRESSO \citep{pepe2021}, EXPRES \citep{jurgenson2016b}, KPF \citep{gibson2016}, and MAROON-X \citep{seifahrt2016b}, and atmospheric facilities such as JWST, are shared resources with broad science programs, so either strategy carries substantial opportunity cost. Additional RV and transit observations therefore remain the most direct way to improve an NRPZ classification for favorable targets, while atmospheric observations provide a complementary subsequent test for selected planets whose bulk-density classifications remain ambiguous.

Atmospheric and mass-radius observations do not provide directly interchangeable measurements, so there is no universal conversion from atmospheric visits to an NRPZ precision. For the five planets in our sample with published HST or JWST atmospheric observations (GJ~357~b, LTT~1445~A~b, TOI-836~b, TOI-260~b, and GJ~9827~b), our idealized calculations require 5--31 balanced-quadrature RV measurements (10--62 for uniform phase coverage) and 12--279 effective photometric samples, corresponding to 1--12 transits under the illustrative cadence adopted in Table~\ref{tab:rad_err}. The published atmospheric studies used one or two observing visits per target \citep{Taylor2025,Batalha2026,Alderson2024,Meech2026,Carleo2021}. For example, the TOI-260~b observations comprised two 5.8~hr JWST/NIRSpec visits \citep{Meech2026}, while our idealized calculation for that planet gives 7 balanced-quadrature RV measurements (14 for uniform phase coverage) and 24 effective photometric samples, equivalent to one transit under the illustrative cadence in Table~\ref{tab:rad_err}. These comparisons should not be interpreted as a universal efficiency ratio because the observations probe different wavelengths and atmospheric species and adopt different model assumptions and detection thresholds.

JWST is presently the primary facility for transmission and thermal-emission measurements of the smallest nearby planets, while HST and high-resolution ground-based spectroscopy can constrain selected upper-atmosphere species. JWST measurements of TRAPPIST-1~b and c have excluded thick atmospheres over portions of the tested model space, and recent thermal phase curves provide stronger constraints on heat redistribution \citep{greene2023,zieba2023,Gillon2026}. These constraints remain dependent on atmospheric composition, clouds, surface properties, and signal-to-noise, and an atmospheric non-detection does not by itself demonstrate that a planet is airless. Atmospheric programs also require substantial resources, such as the Rocky Worlds DDT survey, which was designed around approximately 500~hr of JWST time with complementary HST observations \citep{Redfield2024}. We therefore treat atmospheric characterization as a targeted subsequent step rather than as an automatically more efficient or definitive alternative.

\subsection{Limitations of Our Model}

The NRPZ determinations in this paper rely on ExoPlex \citep{unterborn2023}, but other mass-radius models adopt different equations of state and material properties \citep[e.g.,][]{Seager2003,dorn2015,zeng2016a}. These choices can shift, and to a lesser extent change the width of, the NRPZ at a given radius \citep{Schulze2026}. Such shifts may change the classification of planets close to an NRPZ boundary, but they do not alter our general conclusion that substantially smaller mass and radius uncertainties are required to place most intermediate-density planets securely within or outside the zone.

\subsection{Observational Priorities}

\begin{figure*}
    \centering
    \includegraphics[width=\linewidth]{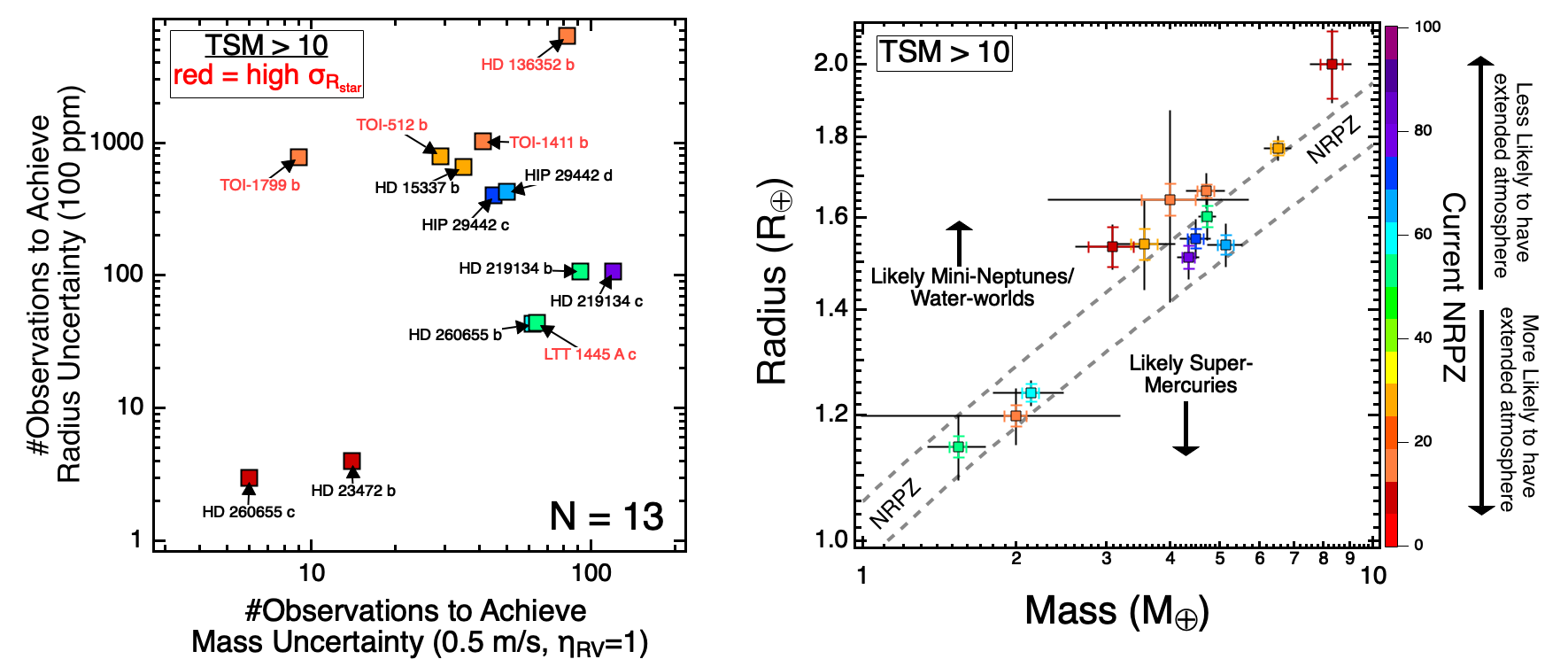}
    \caption{\textbf{Left:} The number of observations needed to lower $\sigma_M$ (assuming 0.5~m~s$^{-1}$ precision and $\eta_{\rm RV}=1$) and $\sigma_R$ (assuming 100~ppm precision) in order to become classified. Those points highlighted in red require lower uncertainties in the stellar radius to achieve a low enough uncertainty in planet radius. \textbf{Right:} Mass and radius for these same planets. The planet's original error bars are shown as black lines, while colored versions represent our recommended uncertainties for classification. Colors in both plots represent the current NRPZ\% using reported mass and radius values. All data in this plot have TSM values $\geq$10.  }
    \label{fig:time_time}
\end{figure*}

Of the 24 planets in our sample that are classifiable, 13 have JWST Transmission Spectroscopy Metrics \citep[TSM; ][]{kempton2018} above $\sim10$, making them potentially favorable candidates for follow-up atmospheric characterization, subject to detailed exposure-time and atmospheric-model calculations. Figure \ref{fig:time_time} highlights the 13 planets in our sample that have TSM values greater than 10 and have not yet had any measurements of their atmospheric composition. These planets deserve consideration for atmospheric follow-up because of their observability and the substantial improvements in mass and radius precision required for NRPZ classification. Of these 13 planets, 7 have current mass, radius and NRPZ\% that make them likely to host an extended atmosphere (HD 15337 b, HD 23472 b, HD 136352 b, HD 260655 c, TOI-512 b, TOI-1799 b and TOI-1411 b) and thus be classified as mini-Neptunes or water worlds. The remaining 6 planets (HD 219134 b, HD 219134 c, HD 260655 b, HIP 29442 c, HIP 29442 d and LTT 1445 A c) are of higher density, and would be consistent with being purely rocky without a significant atmosphere. Other strong candidates for atmospheric follow-up are those with uncertainties in the stellar radius that are too high to significantly reduce the planet's $\sigma_R$ (HD 136352 b, TOI 512 b, TOI 1411 b, TOI 1799 b, and LTT 1445 A c). While classification of each of these planets is possible via mass-radius methods if uncertainties are significantly reduced, following this path would require significant photometry and EPRV efforts. For selected targets, atmospheric observations may provide complementary evidence about the presence of substantial volatile envelopes, but their efficiency relative to further photometric and EPRV observations is target- and model-dependent.

\subsection{Improving Stellar Properties: The Role of Gaia}
\label{gaia}

A frequent source of uncertainty in planetary radii, in addition to the transit photometry itself, is the stellar radius from which the planetary radius is derived. The transit observable is the planet-to-star radius ratio, $R_p / R_\star$, resulting in any uncertainty in $R_\star$ propagating directly and linearly into $R_p$. For much of the Kepler era, stellar radii carried uncertainties of 25--40\%, derived from broadband photometry and spectroscopic $T_\mathrm{eff}$--$\log g$ relations, and these dominated the planet radius error budgets \citep{berger2018a} and habitable zone boundaries \citep{kane2014a}.

The ESA Gaia mission provides precise and accurate parallaxes, combined with broadband photometry and spectroscopic effective temperatures, enabling the determination of stellar luminosities and stellar radii with substantially reduced uncertainties \citep{brown2018}. \citet{berger2018c} used Gaia DR2 parallaxes to revise stellar radii for the Kepler field, reducing typical uncertainties by up to a factor of 10 compared to pre-Gaia values. These Gaia-based revisions altered the derived planetary radii and the inferred planet population. Separately, \citet{vaneylen2018b} used asteroseismic stellar parameters to reveal a sharply defined radius valley.

Gaia DR3 further refined parallaxes and provided the first space-based spectroscopic survey via the Radial Velocity Spectrometer (RVS), delivering homogeneous effective temperatures, surface gravities, and metallicities for millions of stars \citep{brown2021,lindegren2021a,lindegren2021b}. The DR3 astrometric solution achieves median parallax uncertainties of 0.02--0.03~mas for bright sources ($G < 15$), increasing to $\sim$0.07~mas at $G = 17$ and $\sim$0.5~mas at $G = 20$ \citep{lindegren2021c}. For the exoplanet host stars in our sample, the majority of which lie within 50~pc and have $G < 15$, this translates to fractional distance uncertainties of $\sigma_d / d \lesssim 0.1\%$, far below the level at which distance errors contribute meaningfully to the stellar radius error budget. The dominant systematic floor for stellar radii is instead set by the $T_\mathrm{eff}$ scale calibration \citep{stassun2017,stassun2018a} and the accuracy of bolometric corrections, which together contribute $\sim$2--4\% to $R_\star$ even for well-characterized FGK dwarfs \citep{berger2020a}. For M~dwarfs, where empirical $T_\mathrm{eff}$ calibrations carry larger systematics ($\sim$3--5\%), the stellar radius floor is correspondingly higher at $\sim$4--5\%, but still well below the pre-Gaia era uncertainties of $\sim$20--40\%. The forthcoming Gaia DR4, will provide parallax uncertainties approximately $\sqrt{2}$ smaller than DR3 for typical stars, with even larger improvements at the bright end ($G < 13$) where better control over systematics is expected \citep{lindegren2021a}. Gaia DR4 will also include improved treatment of systematic parallax offsets that have been identified in DR3 at the $\sim$10--20~$\mu$as level \citep{lindegren2021b}. The reduction of these systematics will be particularly important for M-dwarf exoplanet host stars, as M dwarfs at distances of 10–40 pc lie precisely in the regime where parallax systematics and photometric calibration uncertainties contribute comparably to $R_\star$ errors.

The practical implication for our candidates is that improved Gaia data releases, particularly DR4, will systematically reduce the $\sigma_{R_p} / R_p$ contribution to bulk density uncertainty across the sample without requiring new dedicated observations. For planets currently in the regime where radius and mass contribute comparably to $\sigma_\rho / \rho$, this improvement in $R_\star$ (and hence $R_p$) may be sufficient to push several planets below the classification threshold when combined with existing or modestly improved mass measurements.

\section{Conclusions}

Here we present results on the uncertainties required to classify a planet as nominally rocky or not-rocky. We adopt the Nominally Rocky Planet Zone (NRPZ) as a proxy for the range of masses and radii that are consistent with a planet being rocky without a significant atmosphere or volatile layer on their surface. We selected a sample of 37 small ($R \leq 2$ R$_\oplus$) planets that orbit bright stars and are most likely to have surface conditions that produce solid, rather than molten, planets. We showed that 12 planets \textit{currently} have mass and radius uncertainties that allow them to be classified as either mini-Neptunes or super-Mercuries. The remaining planets require reductions in their mass and radius uncertainties by an average of $\sim50\%$ and $\sim60\%$ in radius and mass, respectively, for them to be classified as either definitively not rocky, leans not rocky or leans rocky. 

Achieving these improved precisions will require a significant observational investment. Assuming a per-measurement RV precision of 0.5~m~s$^{-1}$ and uniform orbital-phase coverage, we estimate that approximately 10--8100 additional RV measurements are required for the confirmable not-rocky planets, 20--180 for the leaning-not-rocky planets, and 80--240 for the leaning-rocky planets. For radius measurements, assuming a per-measurement photometric precision of 100~ppm, the required total effective photometric sample counts range from approximately 3--31,000, 18--800, and 40--13,600 for the same three groups, respectively. The calendar time required will depend on cadence, orbital-phase coverage, transit sampling, seasonal visibility, and telescope allocation. These estimates are optimistic lower limits because stellar activity, correlated photometric noise, instrumental systematics, telluric contamination, and covariances with additional planets and/or parameters may prevent the uncertainties from decreasing according to the ideal $N^{-1/2}$ relation. Furthermore, the vast majority of confirmed exoplanets orbit relatively faint host stars compared to those in our sample with $V_{mag} < 11$, which intensifies the requirements for sufficient data to adequately reduce the mass and radius uncertainties. For 14 of the 25 planets that require improved radius measurements, the current stellar-radius uncertainty alone exceeds the desired planetary-radius precision, such that improved stellar characterization is required before additional transit photometry can achieve the classification threshold.

These observational requirements raise the question of whether increasingly precise mass and radius measurements provide the most efficient path toward classifying individual terrestrial-sized planets. Bulk density remains valuable for population studies, identifying planets that clearly lie outside the NRPZ, and prioritizing targets for further observations. However, planets lying within or near the NRPZ remain subject to substantial compositional degeneracies even when their masses and radii are measured precisely. Atmospheric characterization provides a complementary and potentially more decisive pathway. The detection of a substantial H$_2$/He or water-rich atmosphere would demonstrate that a planet is not a bare rocky body, whereas sufficiently stringent limits on the presence of a thick atmosphere would increase confidence in a rocky interpretation. Atmospheric non-detections must nevertheless be interpreted cautiously because clouds, hazes, high mean molecular weights, and limited signal-to-noise can suppress observable spectral features. We identify 13 planets in our sample with Transmission Spectroscopy Metric values above the approximate threshold of 10 suggested for terrestrial-planet follow-up with JWST \citep[Figure \ref{fig:time_time},] []{kempton2018}. Because these planets lie within or immediately outside the NRPZ, determining whether they possess substantial atmospheres would both improve their individual classifications and test the broader utility of the NRPZ.

The classification challenge identified here is likely to become more acute as future transit surveys expand the population of terrestrial planets to longer orbital periods. For example, ESA's PLAnetary Transits and Oscillations of stars (PLATO) mission is designed to detect small planets around bright stars, including terrestrial planets in the habitable zones of solar-like stars, while improving their stellar properties through asteroseismology \citep{rauer2025,heller2022b}. The resulting improvements in stellar radii will reduce an important contribution to the planetary-radius uncertainty. Nevertheless, the anticipated PLATO precisions of approximately 5\% in radius and 10\% in mass for an Earth-Sun analogue \citep{rauer2025} remain larger than the approximately 2\% radius and 5\% mass precisions generally required for classification within our NRPZ framework. In addition, the longer orbital periods of many PLATO planets will make further refinement increasingly difficult. Because the RV semi-amplitude scales as $K\propto P^{-1/3}$, the number of RV measurements required to achieve a fixed fractional mass precision scales approximately as $N_{\rm RV}\propto P^{2/3}$ in the white-noise limit, while the observations must also span a substantial fraction of the orbital period \citep[e.g.,][]{cloutier2018b}. Longer-period planets will produce fewer transits during a fixed observing baseline, increasing the calendar time required to refine their radii and ephemerides even though the duration of each individual transit is longer. Long orbital periods will also complicate atmospheric follow-up. Transit and eclipse opportunities for temperate planets may occur only once or twice per year and must coincide with the seasonal visibility windows and scheduling constraints of facilities such as JWST. For periods near one year, successive events recur at nearly the same time of year, so an unfavorable visibility geometry may repeatedly prevent observations. Furthermore, many of the temperate terrestrial planets discovered by PLATO will orbit FGK stars, for which the larger stellar radii produce substantially smaller transmission signals than equivalent planets orbiting late-type M dwarfs \citep{morley2017b,wunderlich2019}. PLATO will therefore greatly enlarge the sample of scientifically compelling terrestrial planets, but may also reveal a population for which precise masses and definitive atmospheric classifications remain observationally expensive or inaccessible. Progress will require combining mass and radius measurements with improved stellar characterization, dynamical information, and atmospheric constraints, while motivating future facilities capable of directly characterizing temperate terrestrial planets.





\section*{Acknowledgements}

The authors acknowledge support from NASA grant 80NSSC23K0267, funded through the NASA Exoplanets Research Program, PI Cayman Unterborn. The results reported herein benefitted from collaborations and/or information exchange within NASA’s Nexus for Exoplanet System Science (NExSS) research coordination network sponsored by NASA’s Science Mission Directorate, grant 80NSSC23K1356, PI Steve Desch. 
This research has made use of the NASA Exoplanet Archive, which is operated by the California Institute of Technology, under contract with the National Aeronautics and Space Administration under the Exoplanet Exploration Program. This research has also made use of the Habitable Zone Gallery at hzgallery.org. The results reported herein benefited from collaborations and/or information exchange within NASA's Nexus for Exoplanet System Science (NExSS) research coordination network sponsored by NASA's Science Mission Directorate.




\bibliographystyle{aasjournal}
\bibliography{references,newrefs}
\section{Supplemental Figures}
\label{sec:Supp}
\begin{figure}
    \centering
    \begin{tabular}{cc}
    \includegraphics[width=0.5\linewidth]{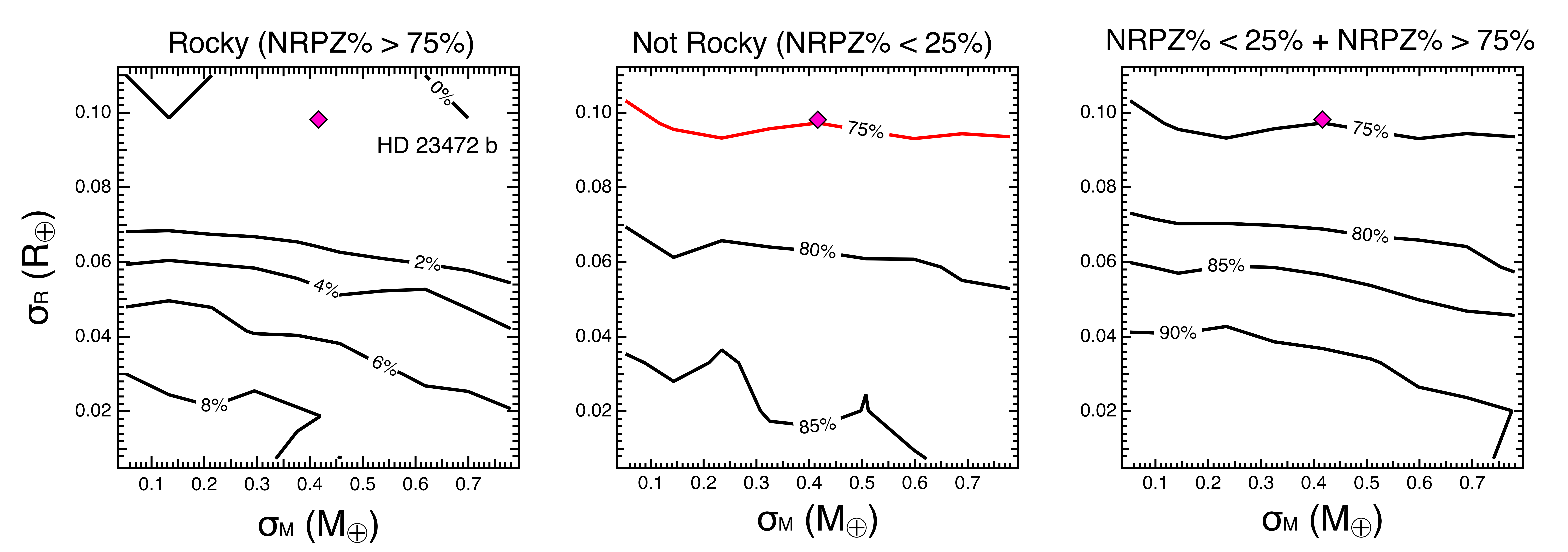} & 
    \includegraphics[width=0.5\linewidth]{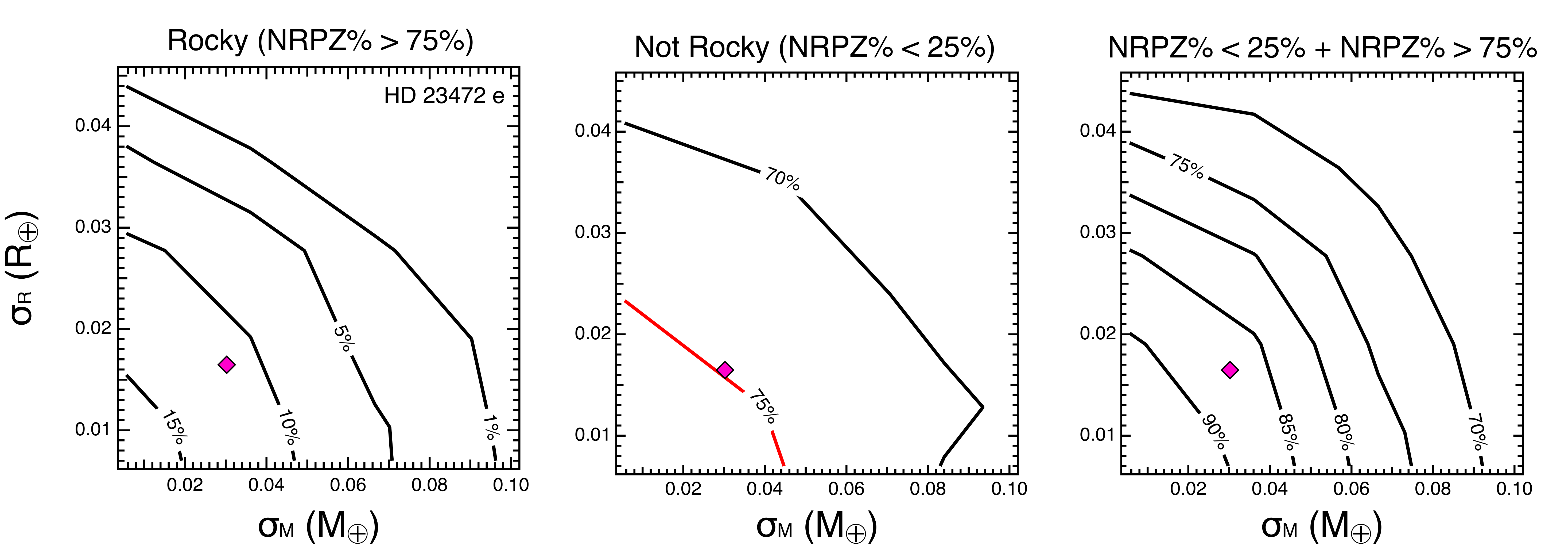}\\
    \includegraphics[width=0.5\linewidth]{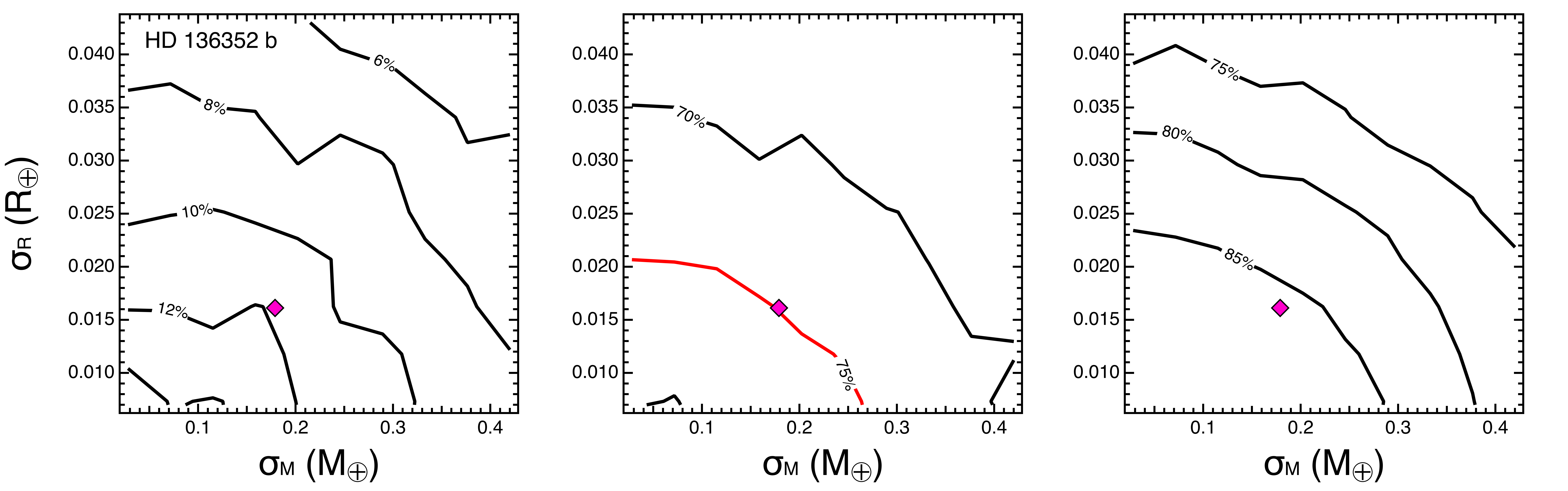} &
    \includegraphics[width=0.5\linewidth]{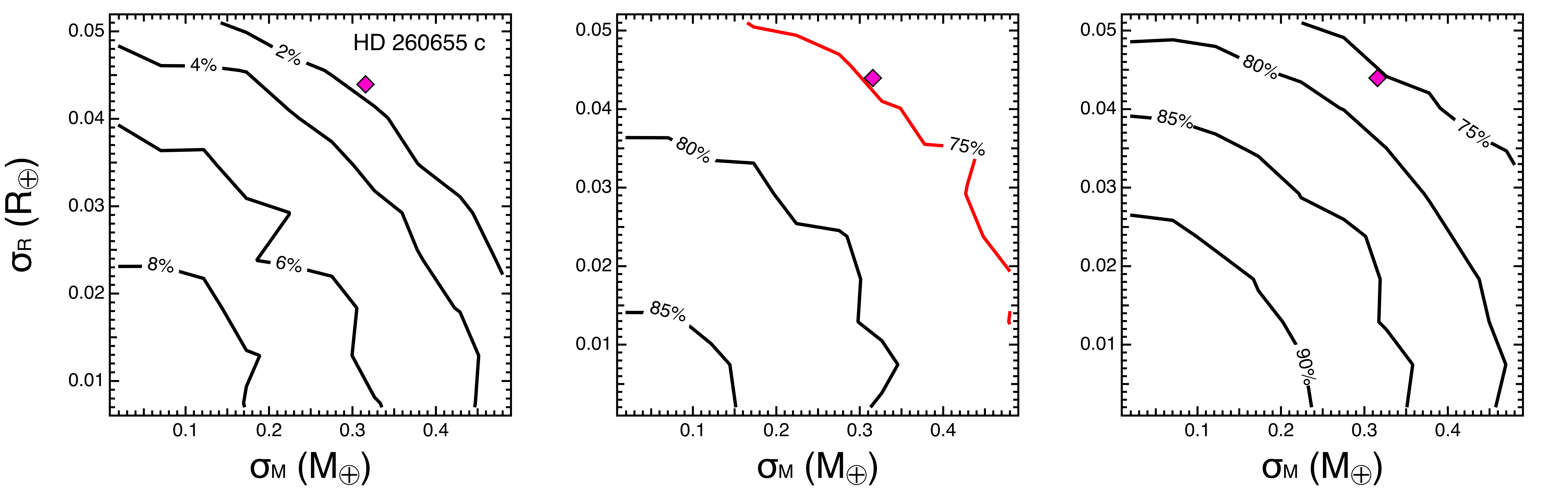}\\
    \includegraphics[width=0.5\linewidth]{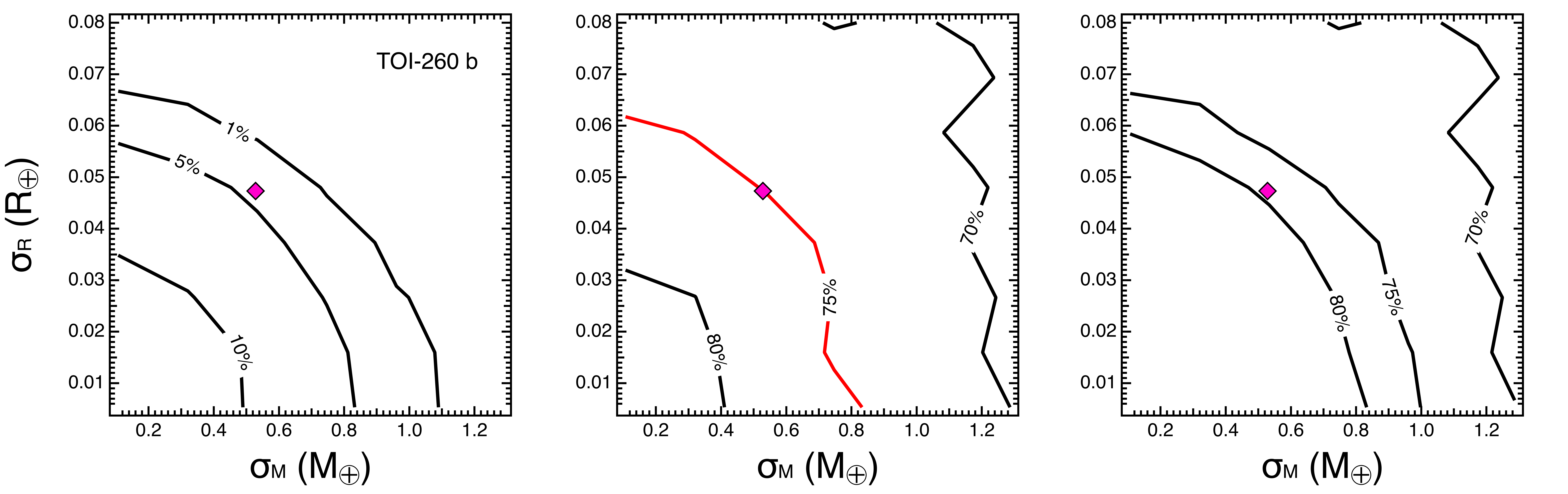} &
    \includegraphics[width=0.5\linewidth]{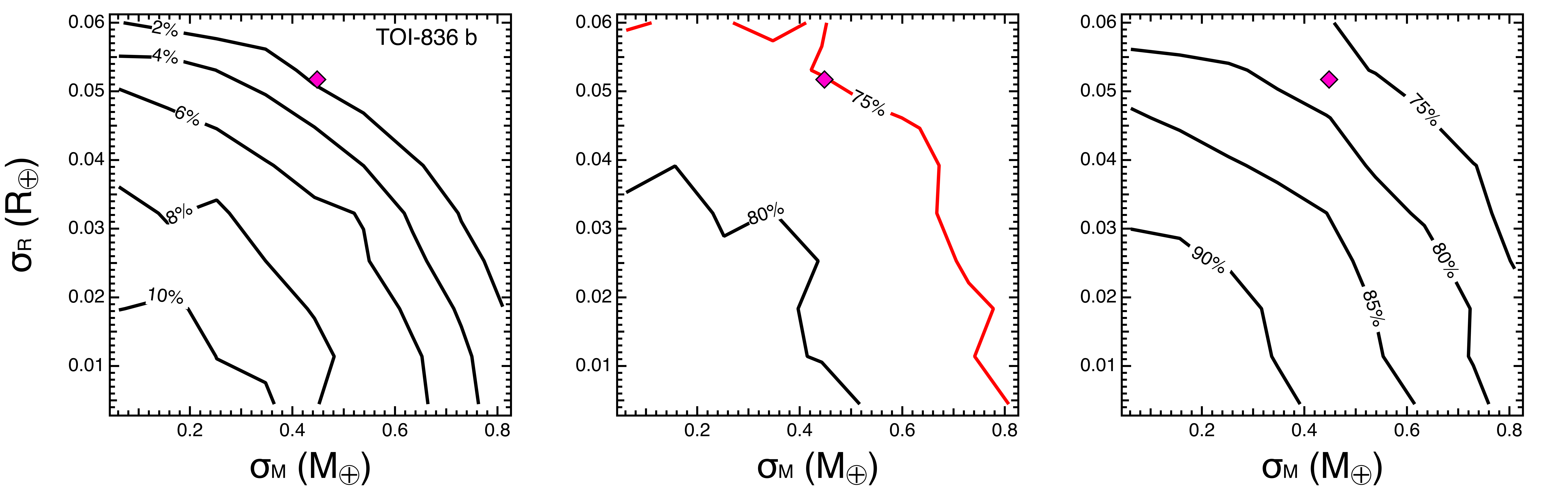}\\
    \includegraphics[width=0.5\linewidth]{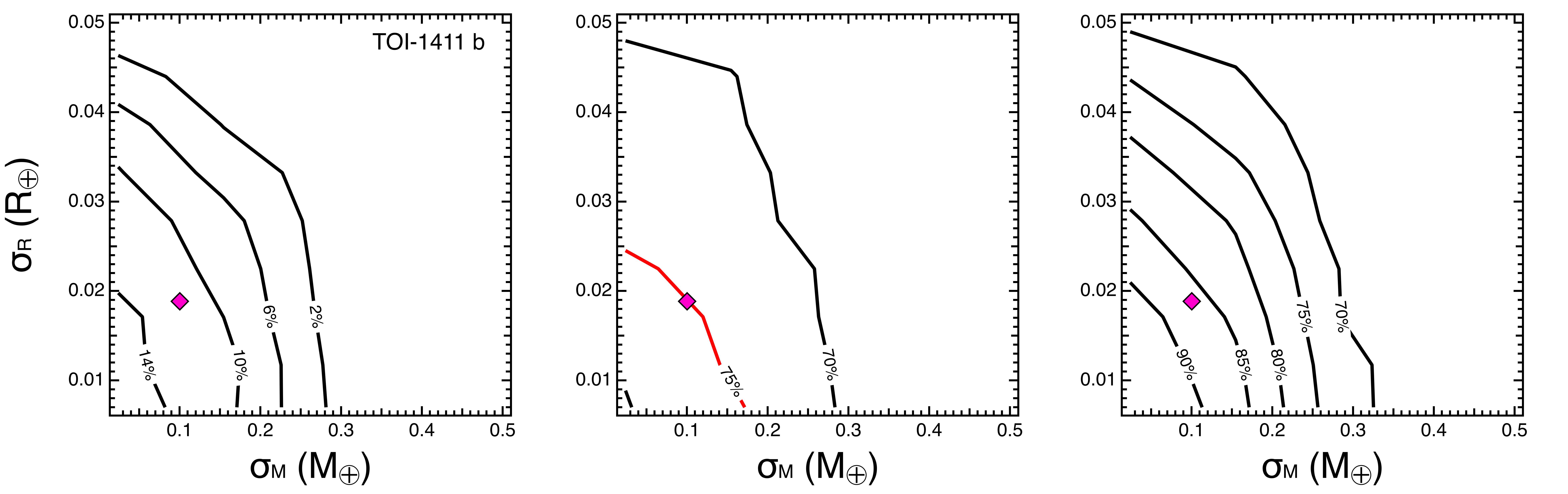}&
    \includegraphics[width=0.5\linewidth]{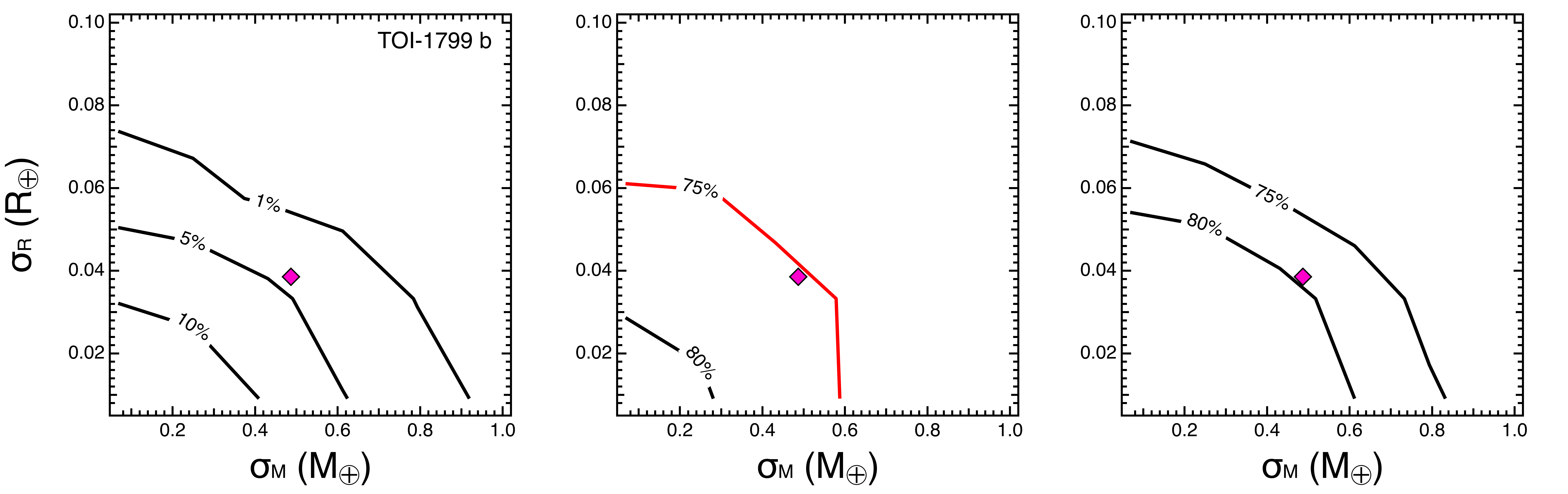}\\
    \multicolumn{2}{c}{\includegraphics[width=0.5\linewidth]{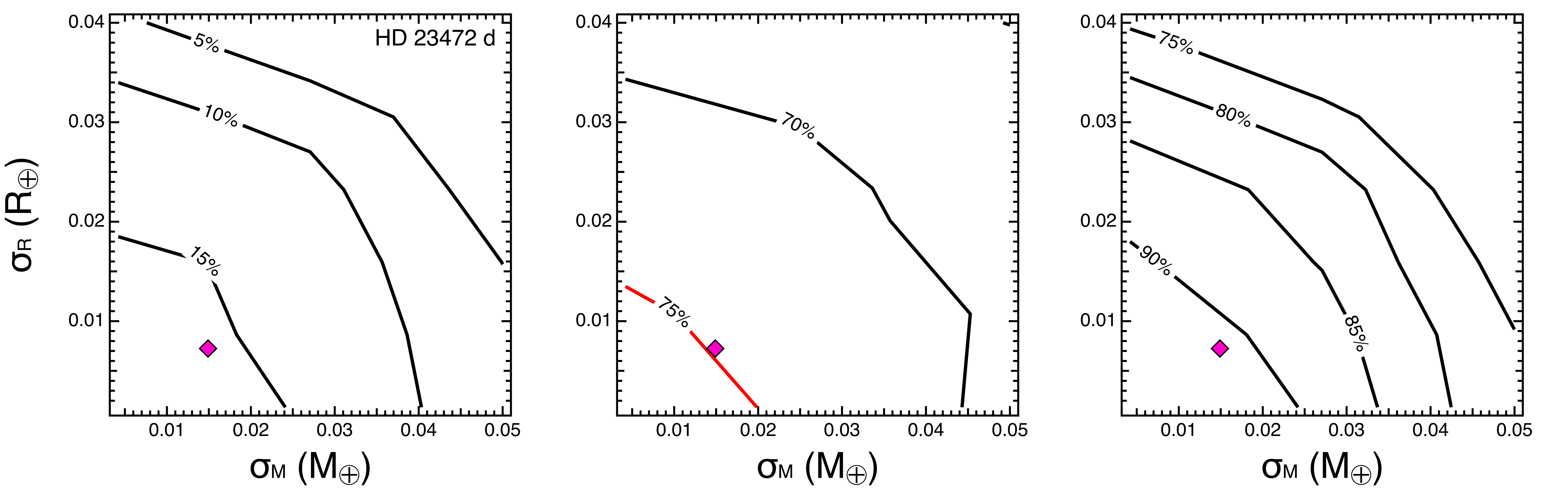}}

    \end{tabular}
        \caption{Contours for each of the confirmable not rocky (CNR) planets in our sample. Best-fit uncertainties in mass and radius are shown as pink diamonds.}
        \label{fig:app_CNR}
\end{figure}

\begin{figure}
    \centering
    \begin{tabular}{cc}
    \includegraphics[width=0.5\linewidth]{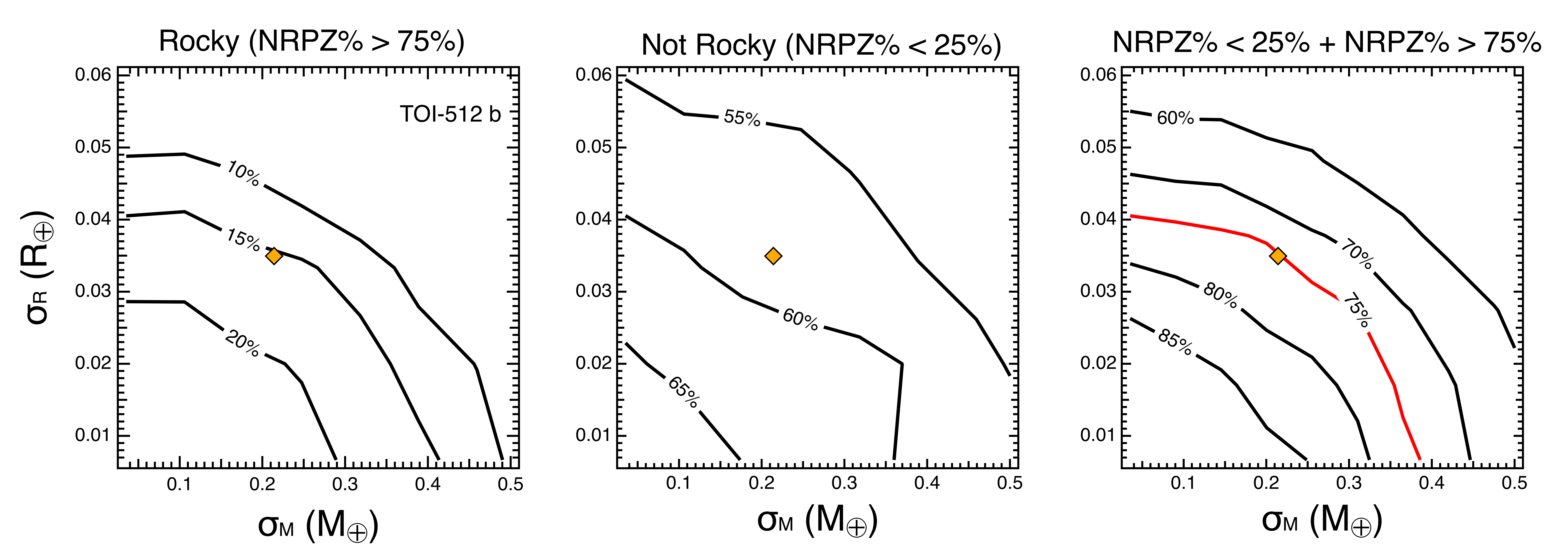} & 
    \includegraphics[width=0.5\linewidth]{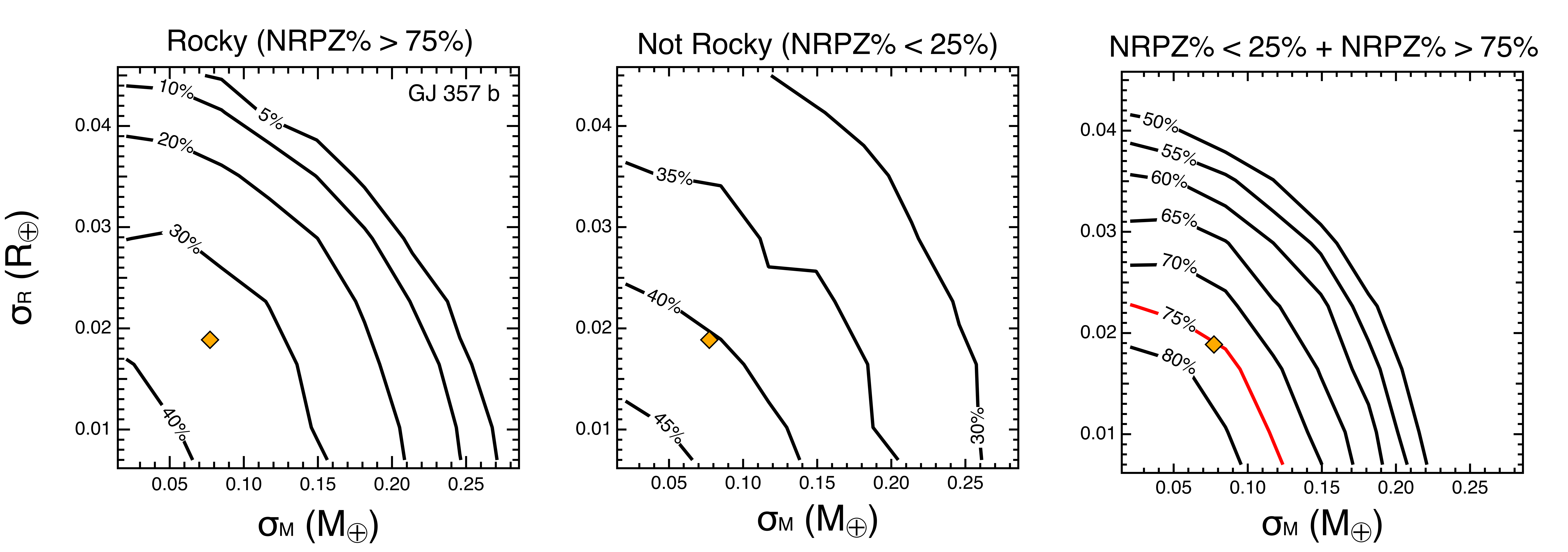}\\
    \includegraphics[width=0.5\linewidth]{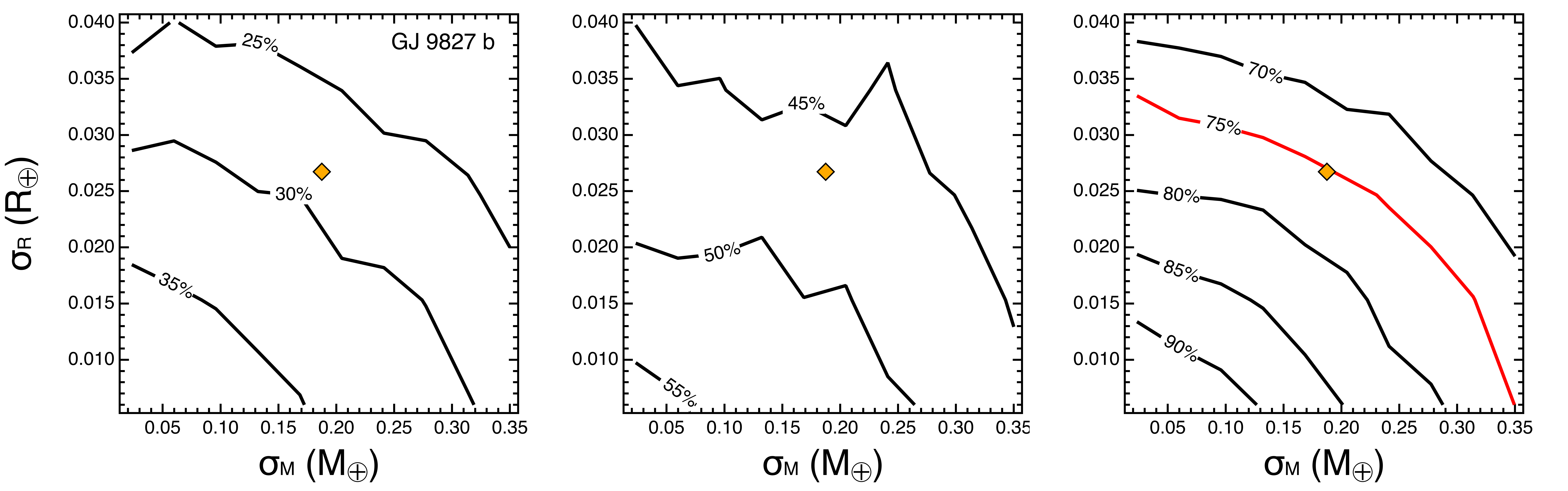} &
    \includegraphics[width=0.5\linewidth]{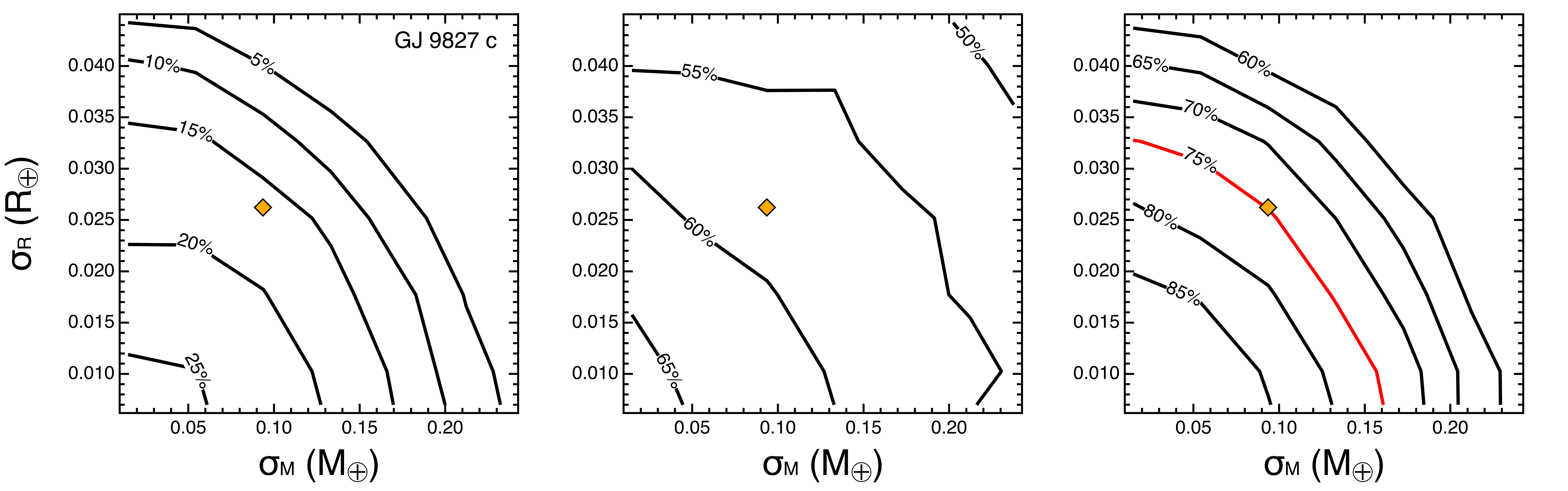}\\
    \includegraphics[width=0.5\linewidth]{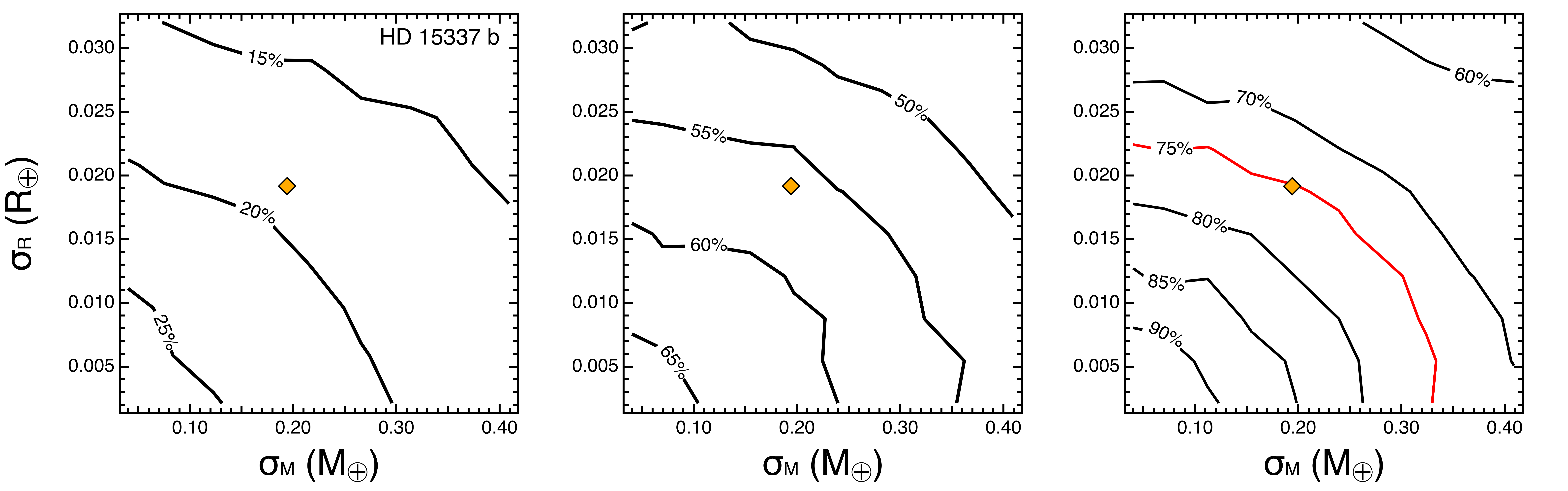} &
    \includegraphics[width=0.5\linewidth]{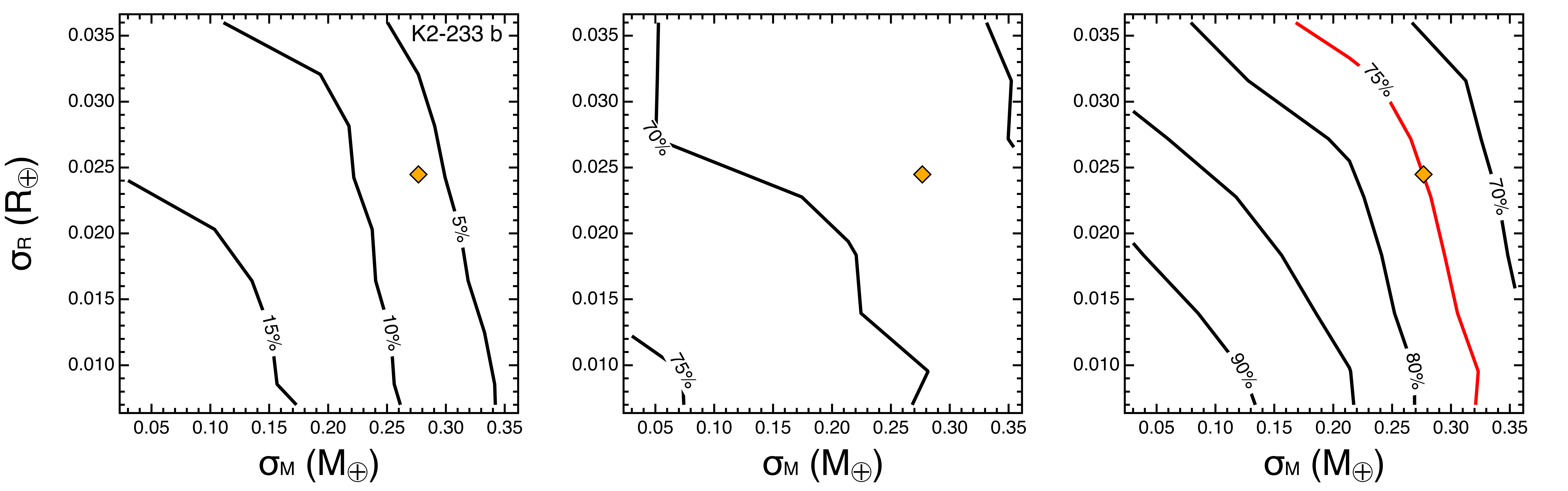}\\
    \includegraphics[width=0.5\linewidth]{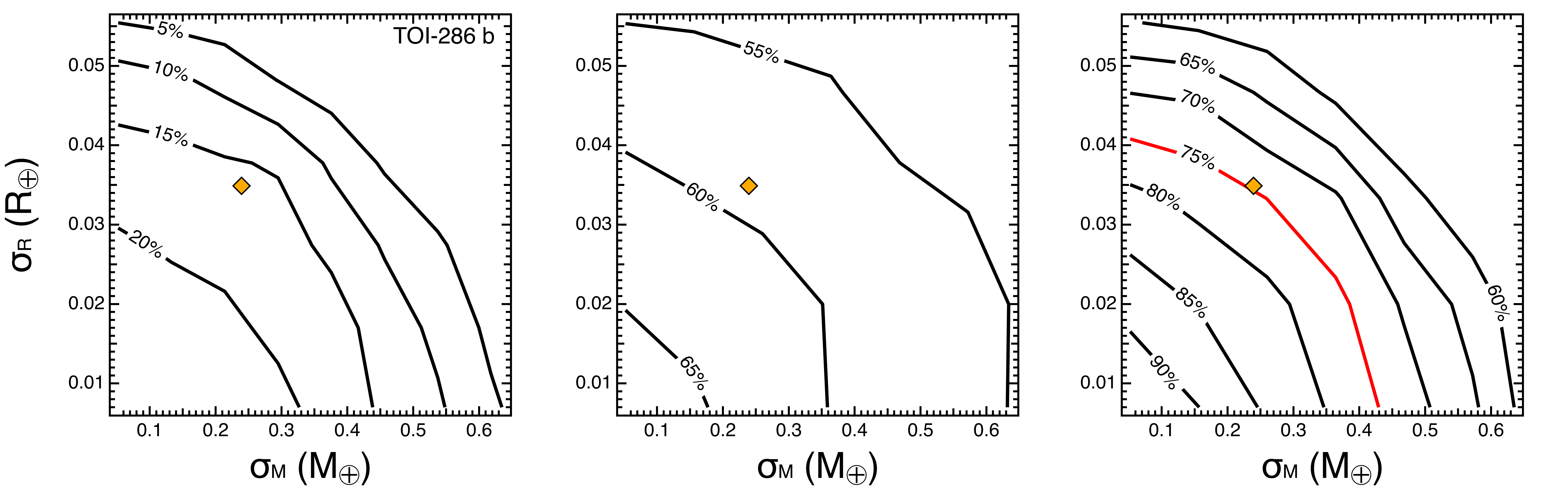}&
    \includegraphics[width=0.5\linewidth]{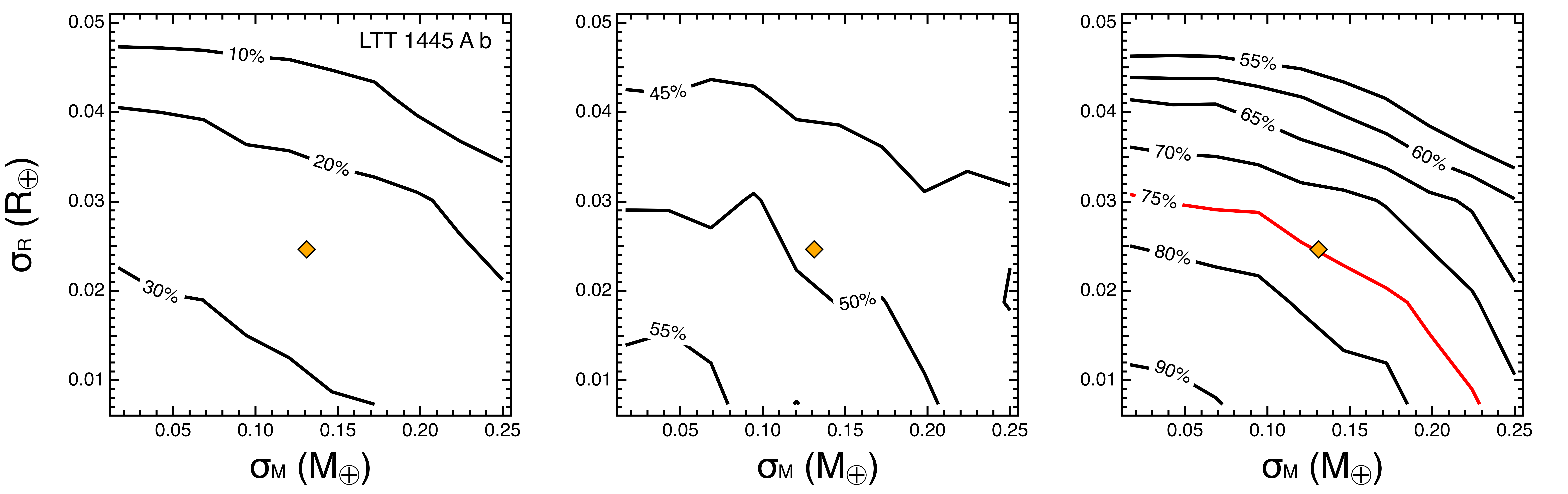}\\

    \end{tabular}
        \caption{Contours for each of the leans not rocky (LNR) planets in our sample. Best-fit uncertainties in mass and radius are shown as orange diamonds.}
        \label{fig:app_NR}

\end{figure}
\begin{figure}
    \centering
    \begin{tabular}{cc}
    \includegraphics[width=0.5\linewidth]{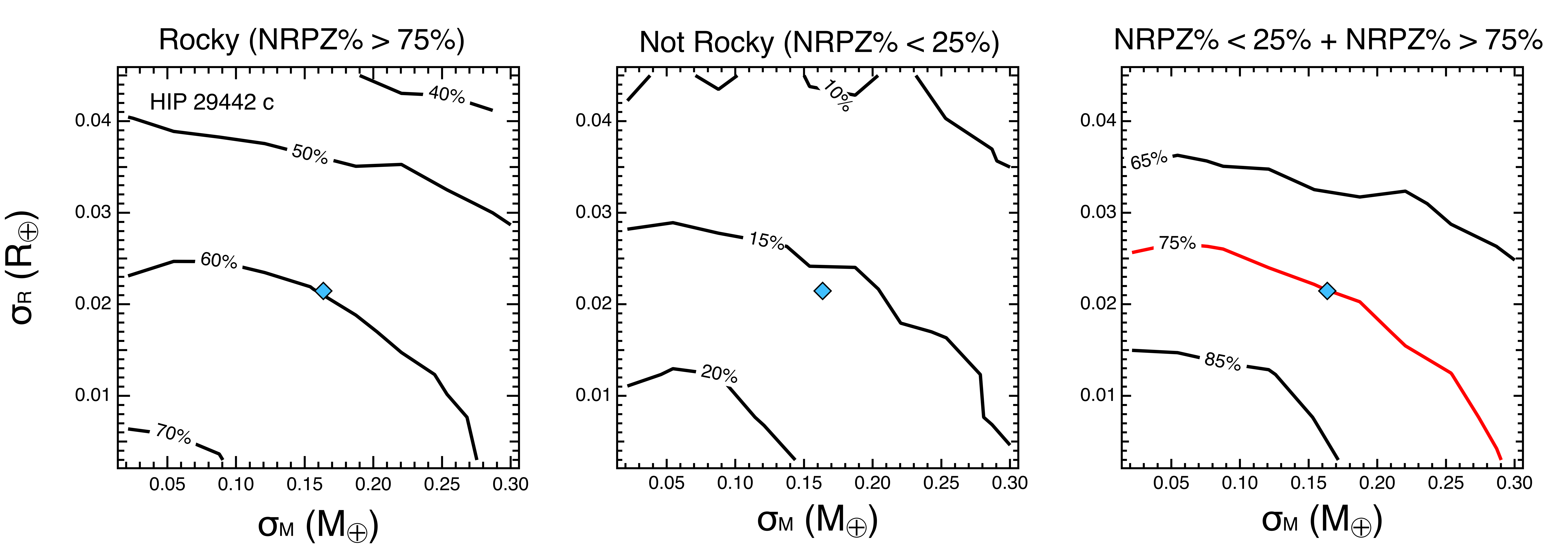} & 
    \includegraphics[width=0.5\linewidth]{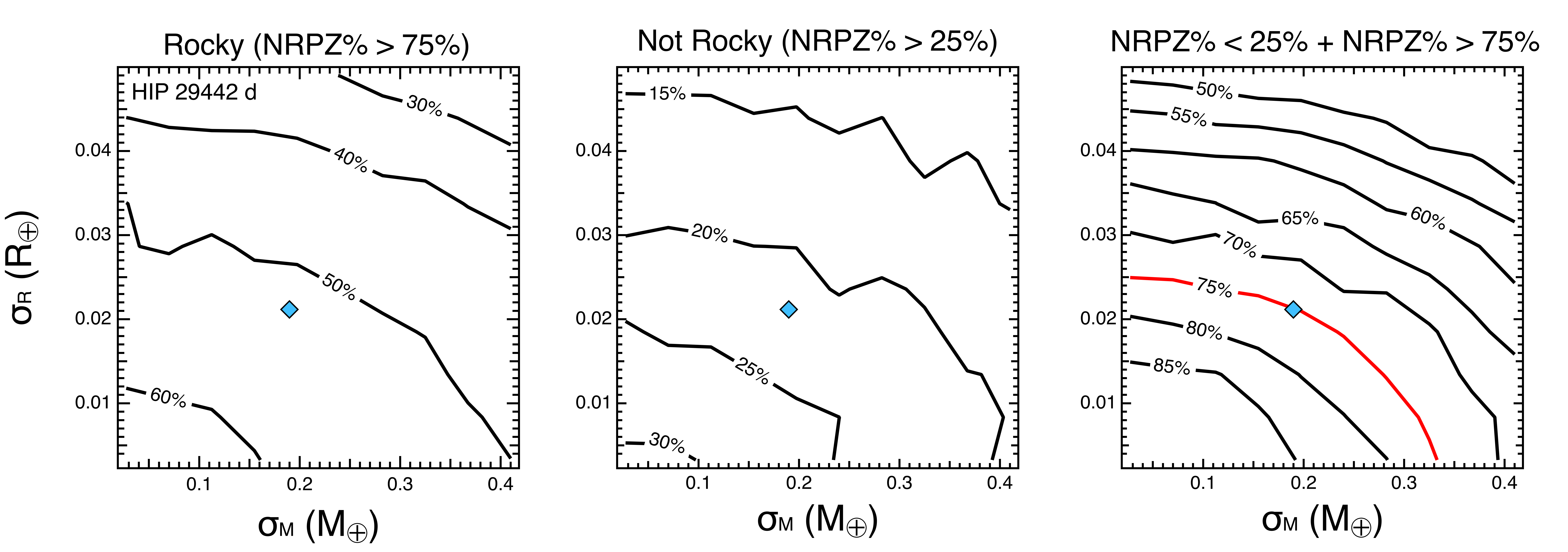}\\
    \includegraphics[width=0.5\linewidth]{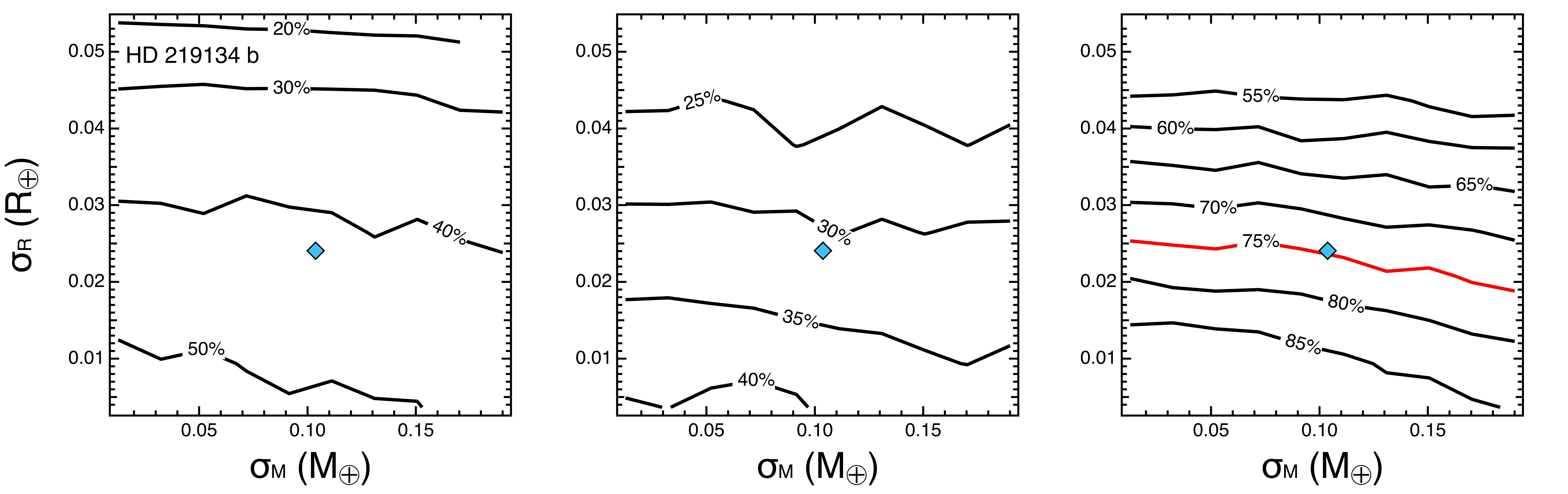} &
    \includegraphics[width=0.5\linewidth]{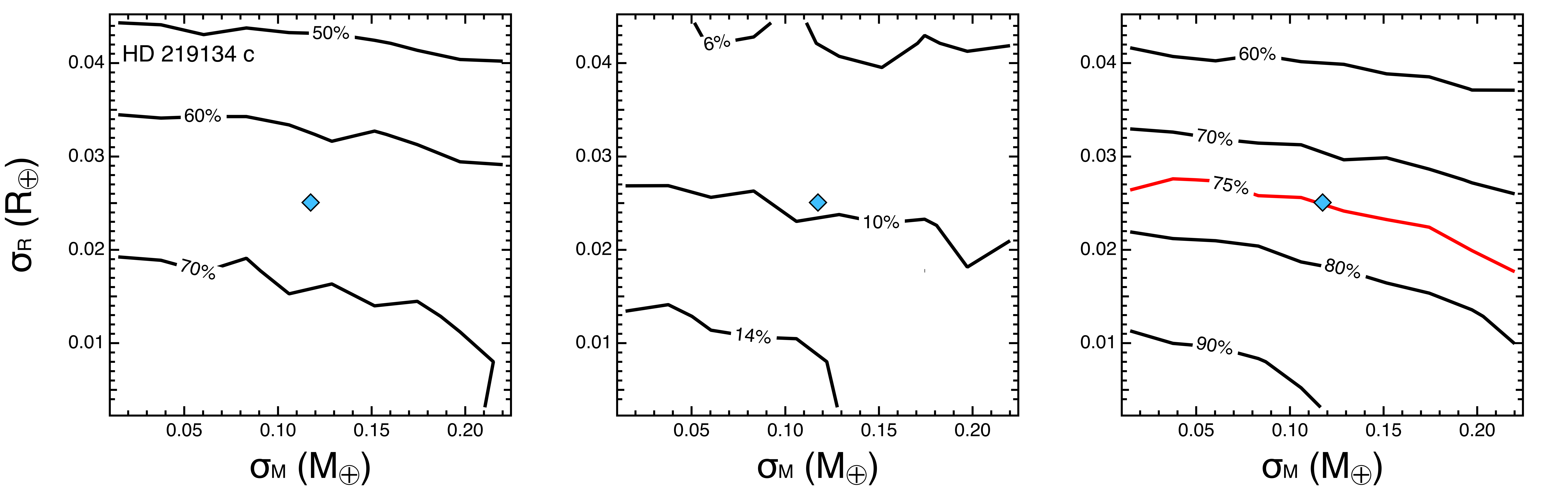}\\
    \includegraphics[width=0.5\linewidth]{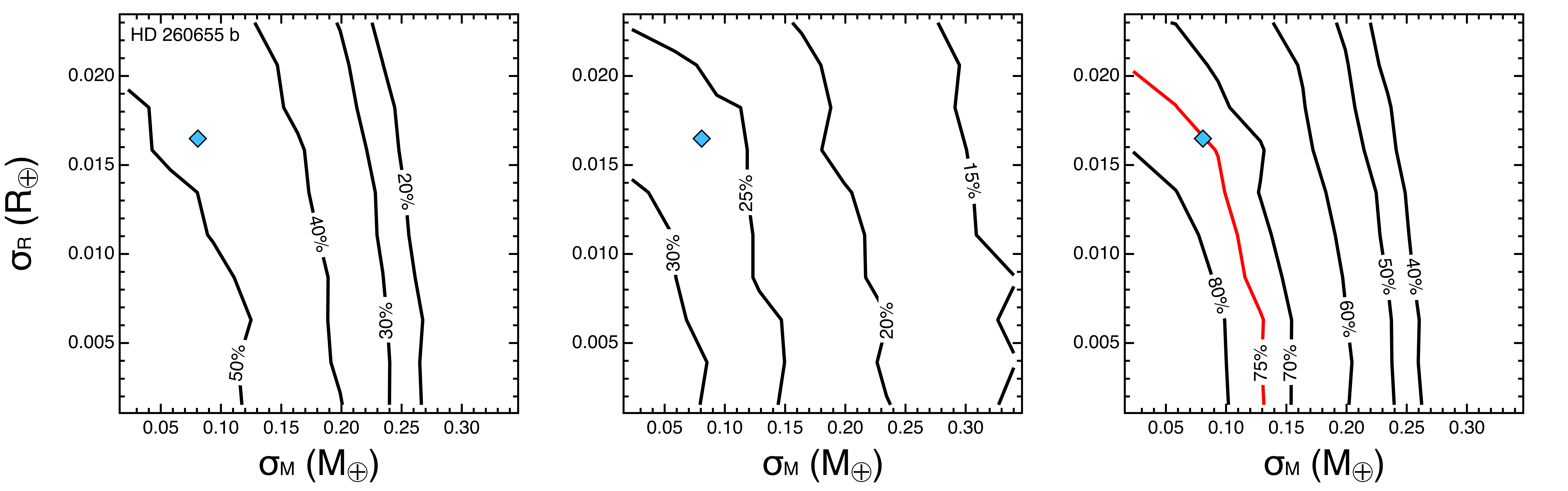} &
    \includegraphics[width=0.5\linewidth]{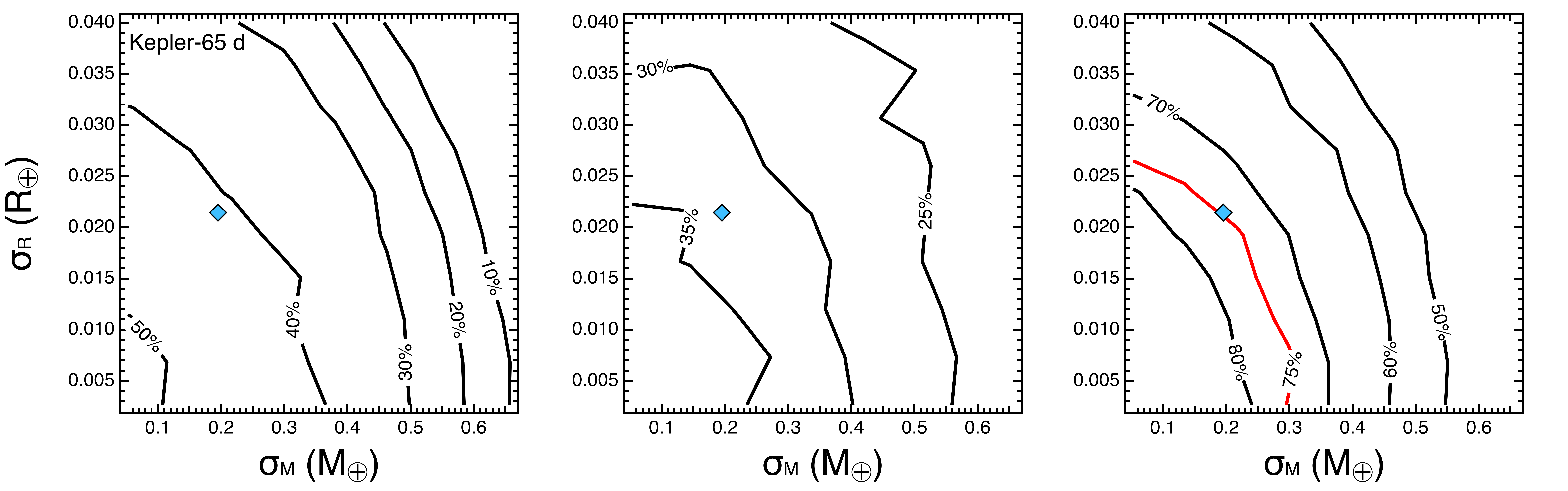}\\
    \includegraphics[width=0.5\linewidth]{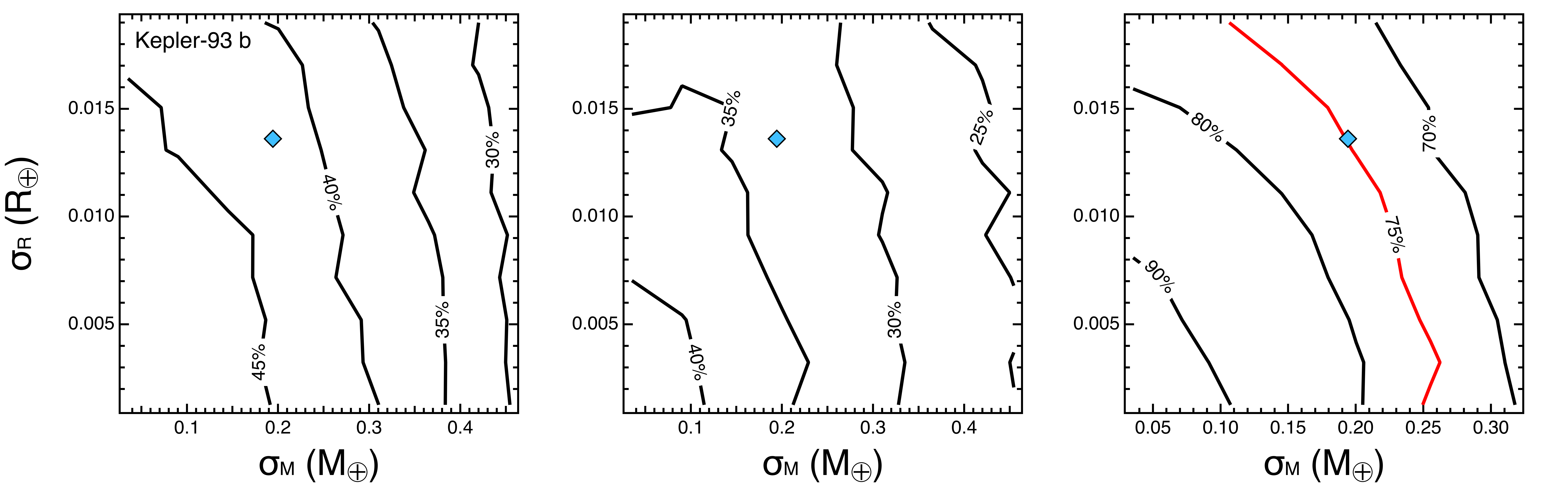}&
    \includegraphics[width=0.5\linewidth]{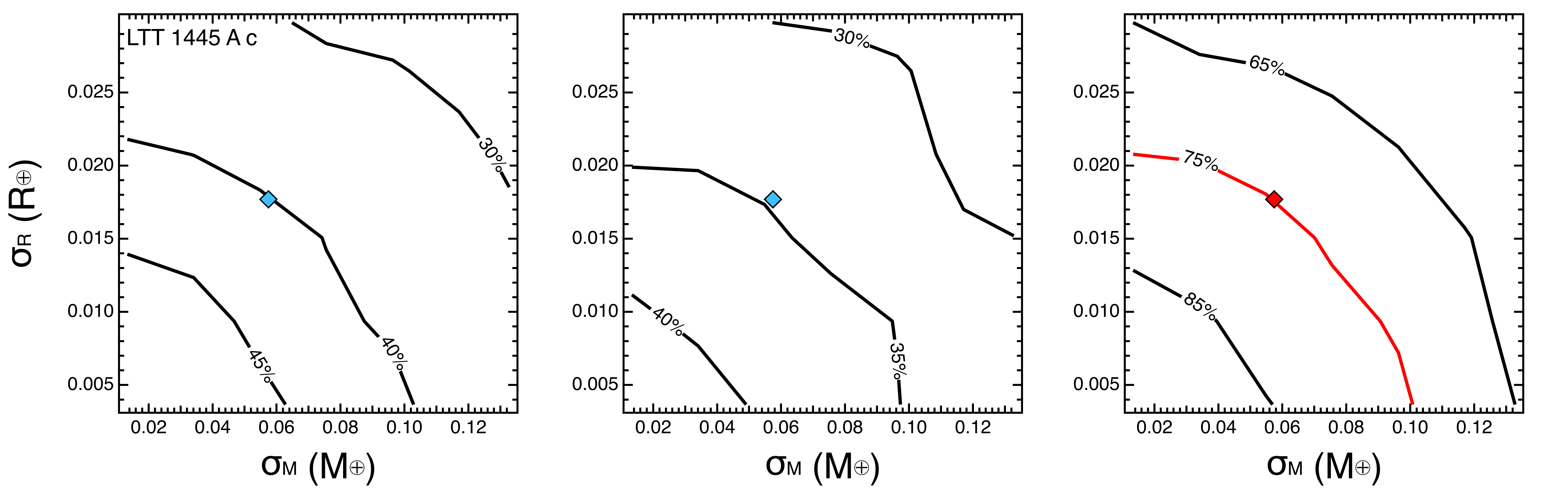}\\

    \end{tabular}
        \caption{Contours for each of the leans rocky (LR) planets in our sample. Best-fit uncertainties in mass and radius are shown as blue diamonds.}
        \label{fig:app_R}

\end{figure}


\end{document}